\documentclass[aip,preprint]{revtex4-1}
\usepackage{graphicx,color}
\usepackage{amsmath}

\begin{document}
\title{
First-principles study of 
intersite magnetic couplings 
\\in
NdFe$_{12}$ and NdFe$_{12}$X (X = B, C, N, O, F)
}
\author{Taro \surname{Fukazawa}}
\email[E-mail: ]{taro.fukazawa@aist.go.jp}
\affiliation{CD-FMat, National Institute of Advanced Industrial Science
and Technology, Tsukuba, Ibaraki 305-8568, Japan}
\affiliation{ESICMM, National Institute for Materials Science,
Tsukuba, Ibaraki 305-0047, Japan}

\author{Hisazumi \surname{Akai}}
\affiliation{The Institute for Solid State Physics, The University of
Tokyo, 5-1-5 Kashiwano-ha, Chiba, Japan}
\affiliation{ESICMM, National Institute for Materials Science,
Tsukuba, Ibaraki 305-0047, Japan}

\author{Yosuke \surname{Harashima}}
\affiliation{CD-FMat, National Institute of Advanced Industrial Science
and Technology, Tsukuba, Ibaraki 305-8568, Japan}
\affiliation{ESICMM, National Institute for Materials Science,
Tsukuba, Ibaraki 305-0047, Japan}

\author{Takashi \surname{Miyake}}
\affiliation{CD-FMat, National Institute of Advanced Industrial Science
and Technology, Tsukuba, Ibaraki 305-8568, Japan}
\affiliation{ESICMM, National Institute for Materials Science,
Tsukuba, Ibaraki 305-0047, Japan}

\date{\today}
\begin{abstract}
We present a first-principles investigation of
 NdFe$_{12}$ and NdFe$_{12}$X (X = B, C, N, O, F)
 crystals with the ThMn$_{12}$ structure.
 Intersite magnetic couplings in these compounds,
 so-called exchange couplings,
are estimated by Liechtenstein's method.
 It is found that the Nd--Fe couplings are sensitive to the 
 interstitial dopant X, with the Nd--Fe(8j) coupling in particular 
 reduced significantly for X = N.
This suggests that
the magnetocrystalline anisotropy 
decays quickly with rising temperature
 in the X = N system although
nitrogenation has advantages over the other dopants in terms of enhancing low-temperature magnetic
properties.
 The Curie temperature
is also calculated from the magnetic couplings
 by using the mean field approximation.
Introduction of X enhances the Curie temperature, with 
 both structural changes and chemical effects found to play important roles
 in this enhancement.

\end{abstract}
\preprint{Ver. 5.0.0}
\maketitle
\section{Introduction}
\label{Introduction}
The magnetic compound NdFe$_{12}$N, which has the ThMn$_{12}$ tetragonal
structure, has attracted much attention
since
Hirayama et al.~\cite{Hirayama15,Hirayama15b}
successfully synthesized the compound---motivated
by the results of a first-principles
calculation~\cite{Miyake14}---and found that
it has higher spontaneous saturation
magnetization
and anisotropy field
than Nd$_2$Fe$_{14}$B.
NdFe$_{12}$ was first synthesized on a substrate, and
then nitrogen was introduced.
The nitrogenation
greatly enhanced
the magnetization and anisotropy field~\cite{Hirayama15};
nitrogenation is known to have this effect
in other similar magnetic compounds also.
Although it has not been reported in the case of NdFe$_{12}$,
typical elements can also enhance the Curie temperature
of rare-earth magnets. 
For example, simultaneous enhancement of
saturation magnetization and the Curie temperature has been 
experimentally observed
in R$_2$Fe$_{17}$C$_x$(Ref. \onlinecite{Sun90,Coey90}),
R$_2$Fe$_{17}$N$_x$(Ref. \onlinecite{Coey90}), and
RTiFe$_{11}$N$_x$(Ref. \onlinecite{Yang91,Yang91b}),
where R denotes a series of rare-earth elements.

The advantages of typical elements in magnets
have also been studied theoretically. 
Kanamori~\cite{Kanamori90,Kanamori06} discussed mixing of
the $d$-states of transition metal sites with the states of 
neighboring atoms in compounds,
and he suggested that the strong ferromagnetism in Nd$_2$Fe$_{14}$B
is possibly attributable to mixing between
the boron elements and the neighboring iron elements. 
The importance of this chemical effect has been 
confirmed by first-principles calculation of light elements 
in iron lattices~\cite{Akai95}.
In a recent first-principles study~\cite{Harashima15e}
on the interstitial X in NdTiFe$_{11}$X (X = B, C, N, O, F),
enhancement of magnetization by the magnetovolume effect and chemical effect
was investigated.

Finite-temperature magnetism in hard-magnet compounds
has also been studied.
It was demonstrated
in a study~\cite{Matsumoto16} of an ab initio spin model
of NdFe$_{12}$N
that 
intersite magnetic coupling between
rare-earth sites (R) and
transition metal sites (T) is a key factor, and
magnetic anisotropy above room temperature is expected to be enhanced significantly 
by strengthening the R--T couplings. 

In this paper, we investigate magnetic couplings in NdFe$_{12}$ and
NdFe$_{12}$X
 for X = B, C, N, O, F based on first-principles calculation. 
The theoretical framework and the computational methods are described in
Section \ref{framework}.
We analyze
the effects of 
the interstitial dopant X on
the magnetic couplings by using
Liechtenstein's
formula.
The obtained intersite magnetic couplings are converted 
to a Curie temperature
by using the mean field approximation.
The dependence of the R--T magnetic couplings on X is 
discussed in Section \ref{SubSecJRT}, and
the values of the Curie temperature 
are discussed in Section \ref{SubSecTc}.

\section{Theoretical Framework and Methods}
\label{framework}
We use AkaiKKR~\cite{AkaiKKR}---a program based
on the Korringa--Kohn--Rostoker (KKR)~\cite{Korringa47,Kohn54}
Green's function method,
which is also known as MACHIKANEYAMA---for calculation of magnetic moments and intersite magnetic
couplings.
This calculation is performed based on the local density
approximation~\cite{Hohenberg64,Kohn65}.

Spin--orbit coupling is considered at only the Nd site, with
the f-electrons treated as an open core of trivalent Nd with the
configuration limited by Hund's rule
(the open-core approximation\cite{Jensen91,Richter98,Locht16}).
We 
apply the self-interaction correction scheme proposed by
Perdew and Zunger\cite{Perdew81} to Nd-f orbitals under
the self-consistent scheme.
The contribution of the Nd-f electrons to
values of the magnetic moment
is not included in our results.
The 1s--5s, 2p--4p, 3d--4d orbitals at Nd;
the 1s--2s, 2p--3p orbitals at Fe;
and the 1s orbital at X and
the 2s orbitals at X = F are treated as core states.
We consider
up to d-scattering ($l_{\max} = 2$) in the systems,
and sample
$6\times 6\times 6$ k-points in the full first
Brillouin zone with reduction of computational 
tasks by exploiting the crystal symmetry.
%
A common set of 
muffin-tin radii is used in 
all calculations for
NdFe$_{12}$ and NdFe$_{12}$X (X = B, C, N, O, F)
such that
the domain volume of local potentials perturbed 
in Liechtenstein's method
does not depend on the system.
We place one X atom per formula unit at the 2b site.

We use lattice parameters obtained via
QMAS~\cite{QMAS}, which is a package for first-principles
calculation based on
the projector augmented-wave (PAW) method~\cite{Bloechl94,Kresse99},
within a generalized gradient
approximation.
These values of the lattice parameters are summarized in 
Appendix \ref{lattparams}.
We refer readers to Ref. \onlinecite{Harashima14b}
for 
details of the calculation setup.

To investigate the dependence of magnetic properties on
lattice parameters, we also perform
calculations of hypothetical
NdFe$_{12}$X systems with
the lattice parameters fixed to the values of other systems.
We write 
``A\#B'' to mean
 ``the system having chemical formula A with the lattice parameters of system B.''
 For example, we examine NdFe$_{12}$\#NdFe$_{12}$X to
 investigate the effect of structural changes (lattice expansion, etc.) induced by the introduction
 of X separately from the chemical effects.
 For further simplicity, 
 we use the ordinary
 chemical formulae NdFe$_{12}$ and NdFe$_{12}$X
 to denote systems with the optimal structure 
unless otherwise stated. 

Our definition of intersite magnetic couplings is as follows.
According to Liechtenstein et al.~\cite{Liechtenstein87},
we map energy shifts caused by spin-rotational perturbations
onto $J_{i,j}$ values
in the following classical Heisenberg Hamiltonian $\mathcal{H}$:
\begin{equation}
 \mathcal{H}
  =
  -\sum_i\,\sum_j\,
  J_{i,j}\,{\vec e}_i \cdot {\vec e}_j,
  \label{Hamiltonian}
\end{equation}
where ${\vec e}_i$ is a unit vector taking the direction of the local spin moment at the $i$th site.
It is also possible to determine
a set of $J_{i,j}$ values by comparing the total energies calculated for
several different
magnetic structures (e.g., different anti-ferromagnetic configurations).
However, to obtain
a required number of unique $J_{i,j}$ values
for a fixed magnetic structure, we exploit Liechtenstein's prescription. 
Our Hamiltonian can be transformed formally into the form of 
$\mathcal{H} = -\sum_{i,j} \vec{S}_i \, \mathcal{J}_{i,j} \, \vec{S}_j$,
with
$J_{i,j} = S^0_i{\mathcal J}_{i,j}S^0_j$
and 
${\vec S}_i = S^0_i\,{\vec e}_i$, where $S^0_i$ denotes 
the local moment of the $i$th site in the ground state.
Note, however, that $\mathcal{J}_{i,j}$ is an indirect outcome
from Liechtenstein's scheme, whereas $J_{i,j}$ is more directly related
to the energy shift under the perturbation at the $i$th and $j$th sites
considered in the scheme.
Therefore, we discuss values of $J_{i,j}$
instead of $\mathcal{J}_{i,j}$ to maintain
theoretical clarity.

%



\section{Results and discussion}
\label{discussion}
\subsection{Chemical effects of X on $J_{\rm R\operatorname{--}T}$}
\label{SubSecJRT}

\begin{figure*}
 \includegraphics[width=5.5cm,angle=270]{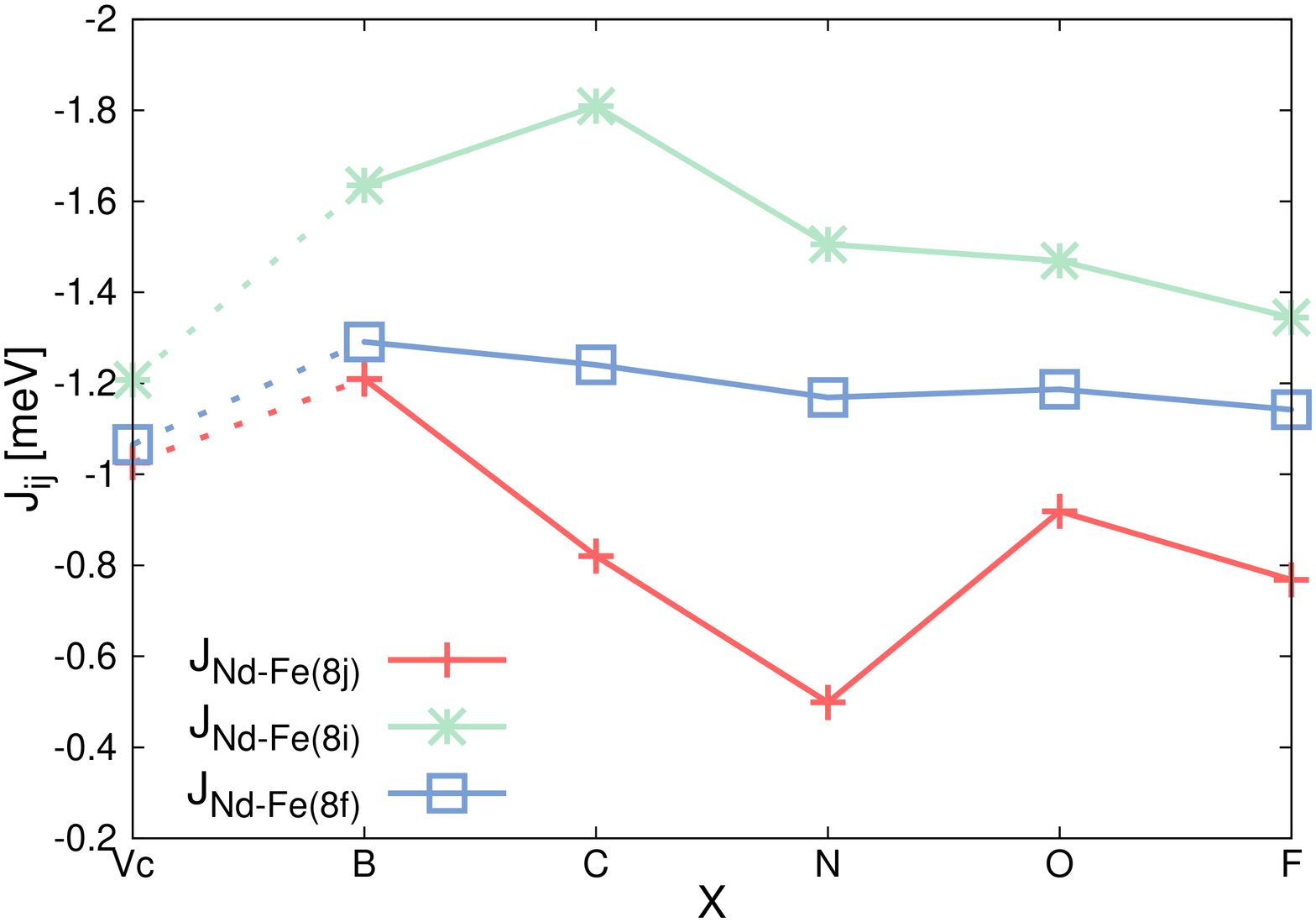}
 \includegraphics[width=5.5cm,angle=270]{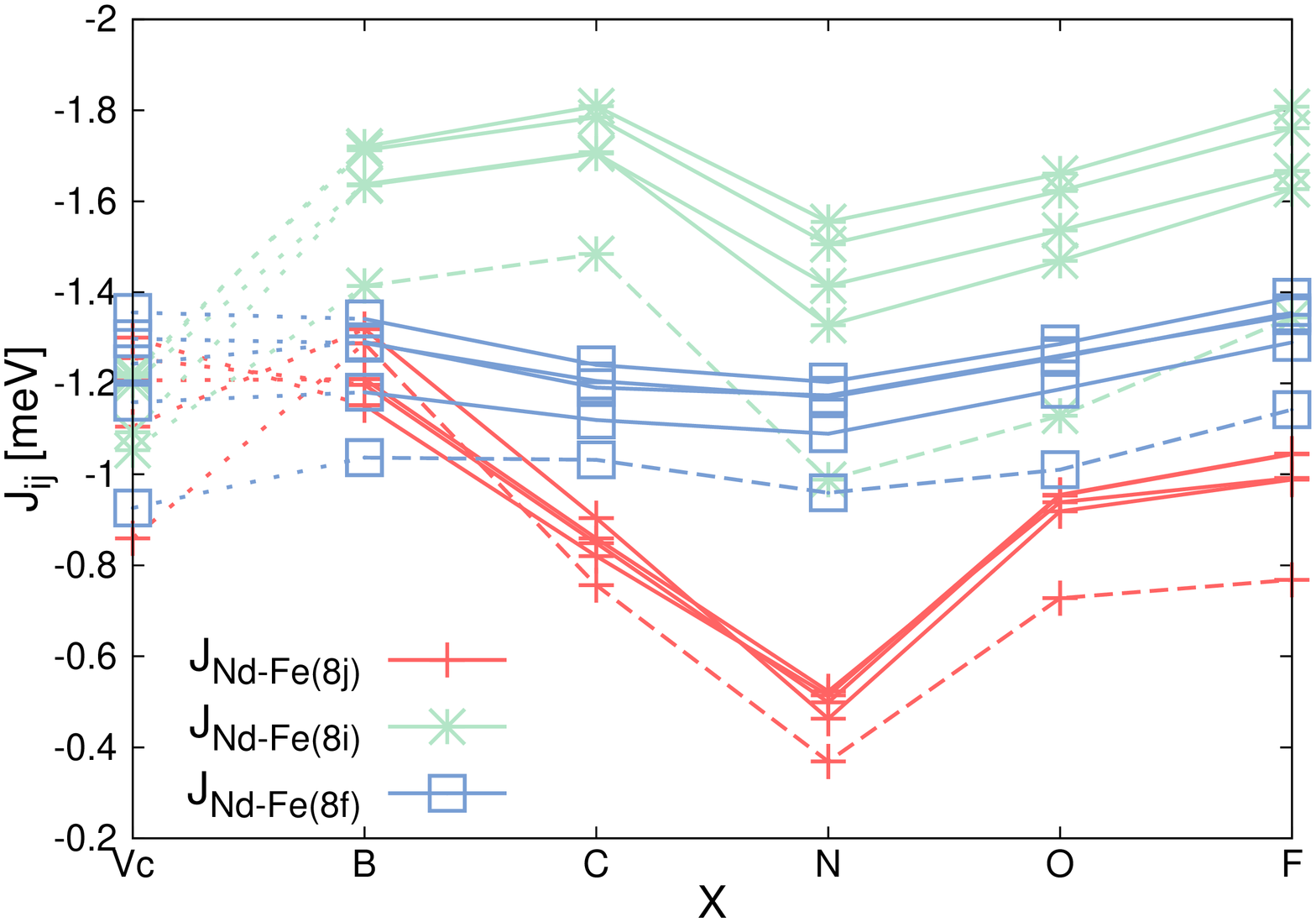}
 \caption{(Color online) 
 (Left) Values of $J_{\rm R\operatorname{--}T}$ for NdFe$_{12}$X  (X = Vc, B, C, N, O, F),
 where NdFe$_{12}$Vc denotes NdFe$_{12}$.
 (Right) Values of $J_{\rm R\operatorname{--}T}$ for NdFe$_{12}$X\#NdFe$_{12}$Z  (Z = B, C, N, O, F).
 Values of 
 $J_{\rm R\operatorname{--}T}$ calculated for the same element Z are connected by lines,
 with dashed lines indicating
 Z = F.
 \label{JRT}
}
\end{figure*}

Figure \ref{JRT} shows the magnetic coupling constants
$J_{\rm R\operatorname{--}T}$ ($J_{i,j}$ of R--T bonds)
for NdFe$_{12}$ (denoted by X = Vc) and 
NdFe$_{12}$X for X = B, C, N, O, F. 
There are three iron sublattices (8j, 8i, and 8f) in the ThMn$_{12}$ structure, and 
values of $J_{\rm R\operatorname{--}T}$ for the shortest Nd--Fe(8j), Nd--Fe(8i), and Nd--Fe(8f) bonds 
are shown.
The three coupling constants have similar values in NdFe$_{12}$. 
The
introduction of X changes the strengths of the couplings significantly,
with the magnitudes of the changes
differing considerably between the different couplings.
The 
$J_{\rm Nd\operatorname{--}Fe(8j)}$ coupling exhibits the strongest
dependency on X, 
which is understandable because Fe(8j) is the closest site to X. 
The $J_{\rm Nd\operatorname{--}Fe(8j)}$ coupling is reduced in all
cases where X is introduced, with the exception of X = B.
The smallest value is found at X = N. That value 
is less than half of NdFe$_{12}$. 
In contrast,
$J_{\rm Nd\operatorname{--}Fe(8i)}$ and
$J_{\rm Nd\operatorname{--}Fe(8f)}$
are enhanced, although slightly, by N.
Nitrogenation thus appears to effectively reduce
the Nd--Fe coupling strengths. 
This suggests that nitrogenation
exacerbates unfavorable thermal dumping of 
magnetocrystalline anisotropy, although it induces strong uniaxial 
anisotropy at low temperatures~\cite{Matsumoto16}.

To examine the chemical effects separately from the structural effects, 
the results for NdFe$_{12}$X\#NdFe$_{12}$Z (Z=B, C, N, O, F) are plotted in 
the right panel of Fig. \ref{JRT}.
It can be seen that the overall trends as a function of X
are not highly dependent on Z.
This indicates that chemical effects dominate this
behavior.
Looking at the results in more detail,
it can be seen that
each of the absolute values of
the $J_{\rm R\operatorname{--}T}$ constants has its minimum at X = N
notwithstanding the choice of lattice parameters.
Furthermore, the reductions caused by N are smaller
than
the fluctuations due to
the structural changes
in the cases of 
$J_{\rm Nd\operatorname{--}Fe(8i)}$ and
$J_{\rm Nd\operatorname{--}Fe(8f)}$.
However, the reduction in $J_{\rm Nd\operatorname{--}Fe(8j)}$ due to N
is too large to be overcome by the structural effect.

\begin{figure}
 \includegraphics[width=5.5cm,angle=270]{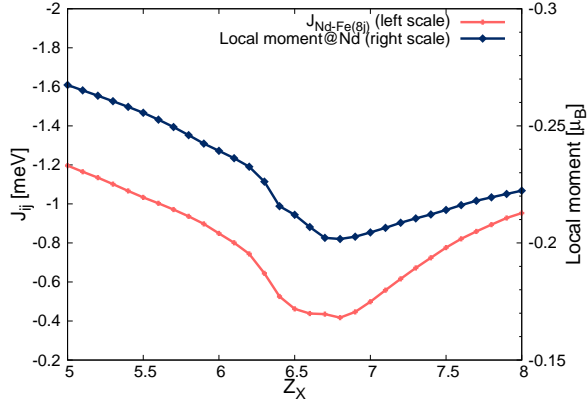}
 \caption{\label{JRT_fixed_lattice}
(Color online) 
 $J_{\rm Nd\operatorname{-}Fe(8j)}$ (left scale)
 for NdFe$_{12}$X\#NdFe$_{12}$N
 as a function of  $Z_{\rm X}$ --- the atomic number of X ---
 compared with the local moment of the Nd site (right scale).
 Note that the right scale for the local moments is inverted.
 }
\end{figure}

Figure \ref{JRT_fixed_lattice} shows
$J_{\rm Nd\operatorname{--}Fe(8j)}$  
for NdFe$_{12}$X\#NdFe$_{12}$N
with fractional changes to the atomic number of X, $Z_{\rm X}$.
This perturbation on $Z_{\rm X}$ serves as a theoretical
probe to the system, and the external potential with
the fractional $Z_{\rm X}$ can also be
interpreted as a model within the virtual crystal approximation
for random occupation of the X site by multiple elements 
with the mean value of their atomic numbers
coinciding with $Z_{\rm X}$.
In Fig. \ref{JRT_fixed_lattice},
there is a significant decrease of the magnitude
at approximately
$Z_{\rm X}$ = 7 (X = N), 
with a minimum at $Z_{\rm X}$ = 6.8.
The figure also shows
the local spin moment at the Nd site
as a function of $Z_{\rm X}$.
This local moment
is strongly correlated with $J_{\rm Nd\operatorname{--}Fe(8j)}$,
so the reduction in $J_{\rm Nd\operatorname{--}Fe(8j)}$ by
nitrogenation
is attributable to the reduction in the spin magnetic moment at
the Nd site.
Note that
$J_{\rm Nd\operatorname{--}Fe(8j)}$
does not necessarily change proportionally to
the local moments because
$\mathcal{J}_{i,j}$ in
$J_{i,j}=
S^0_i\,
\mathcal{J}_{i,j}\,S^0_j$
also depends on the ground state of the system.

\begin{figure}
 \includegraphics[width=5.5cm,angle=270]{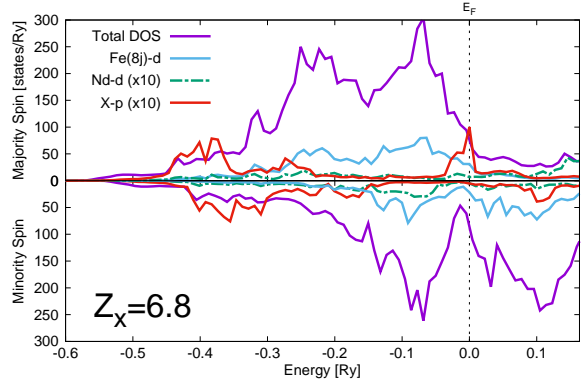}
 \caption{\label{JRT_fixed_lattice_DOS}
(Color online) 
 Total and partial DOS of NdFe$_{12}$X\#NdFe$_{12}$N at 
 $Z_{\rm X}=6.8$.
 Note that the partial densities of the Nd-d states and
 the X-p states are magnified tenfold.
 }
\end{figure}

\begin{figure*}
 \includegraphics[width=5.5cm,angle=270]{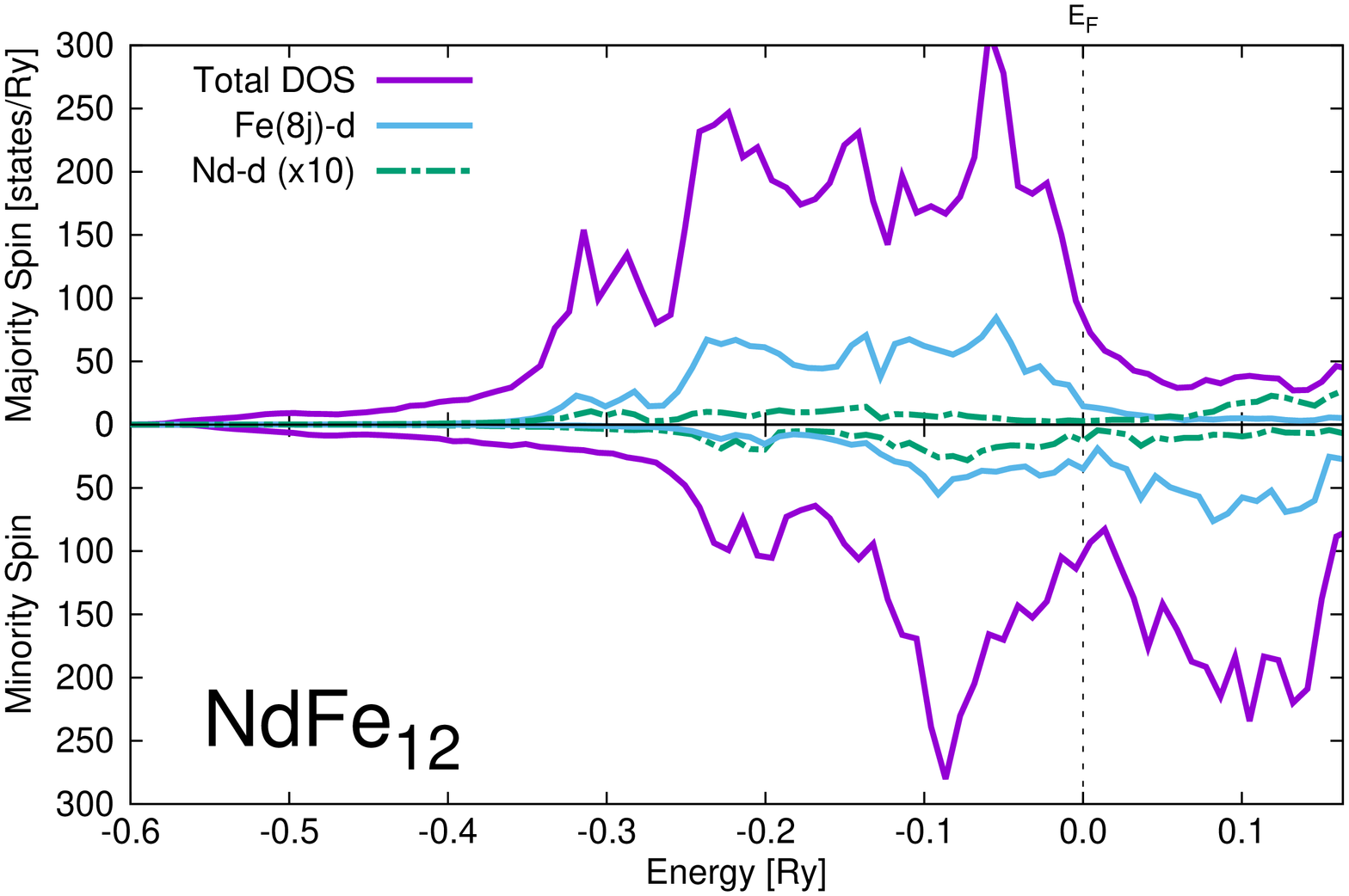}
 \includegraphics[width=5.5cm,angle=270]{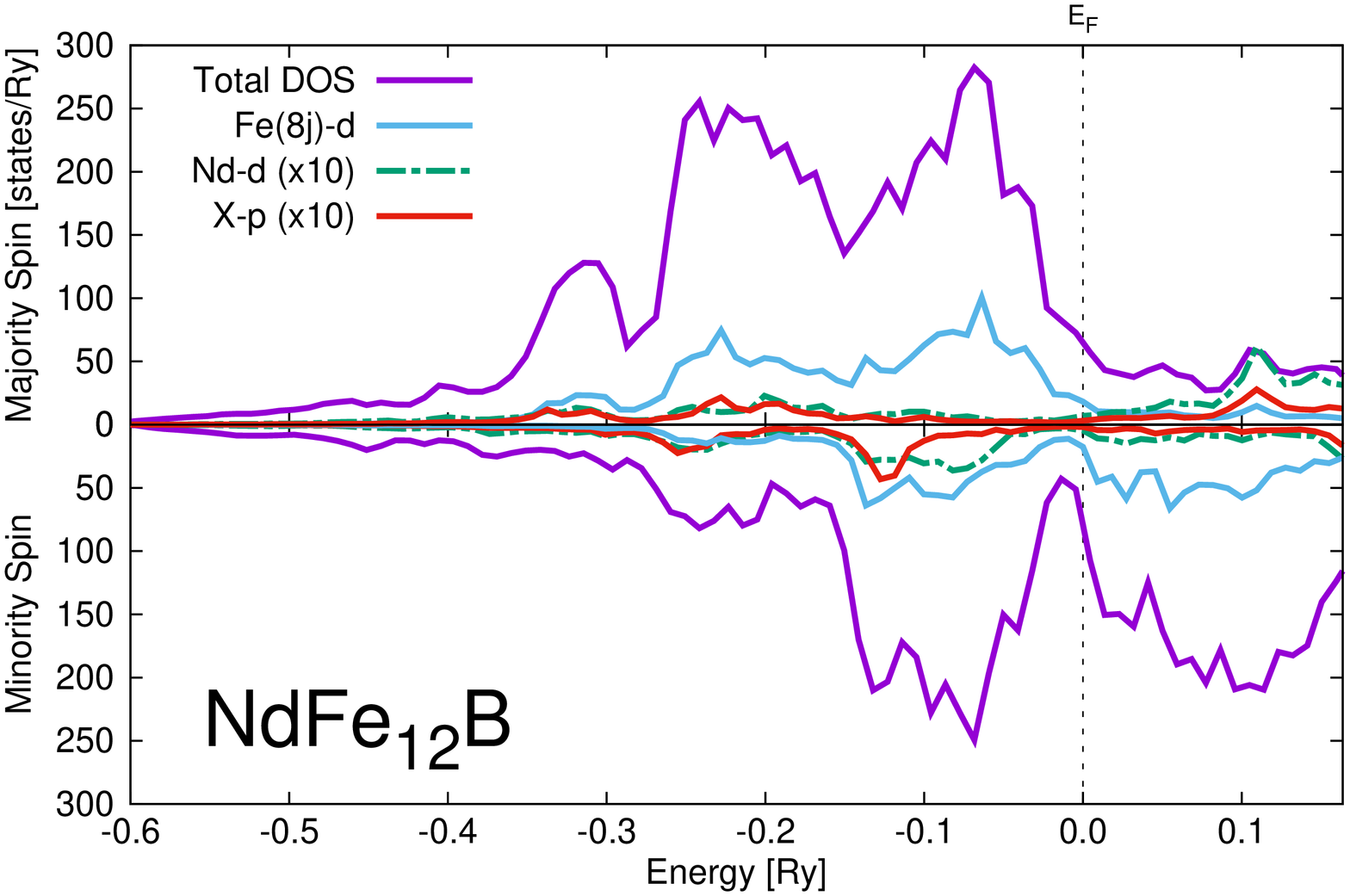}
 \includegraphics[width=5.5cm,angle=270]{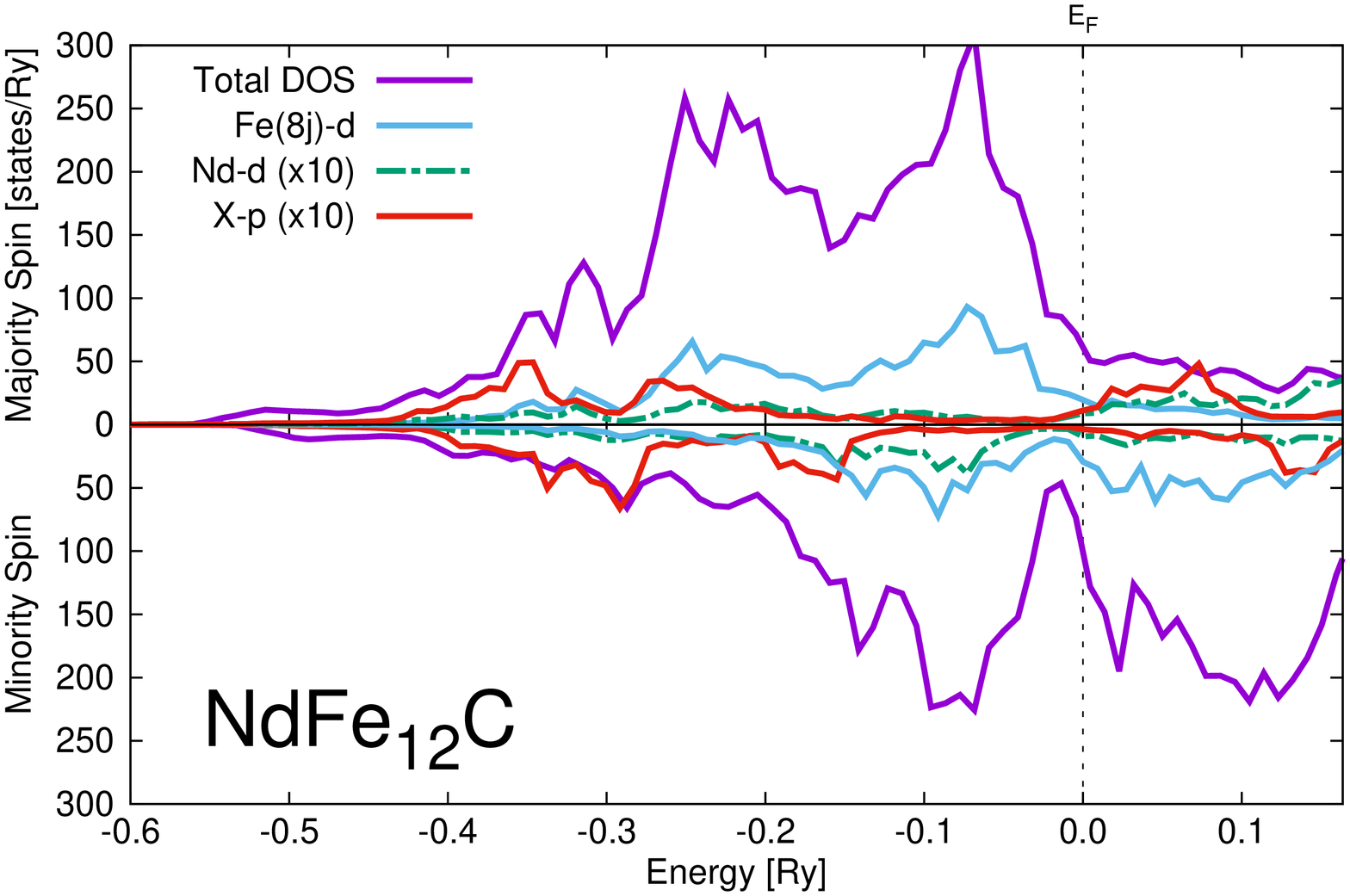}
 \includegraphics[width=5.5cm,angle=270]{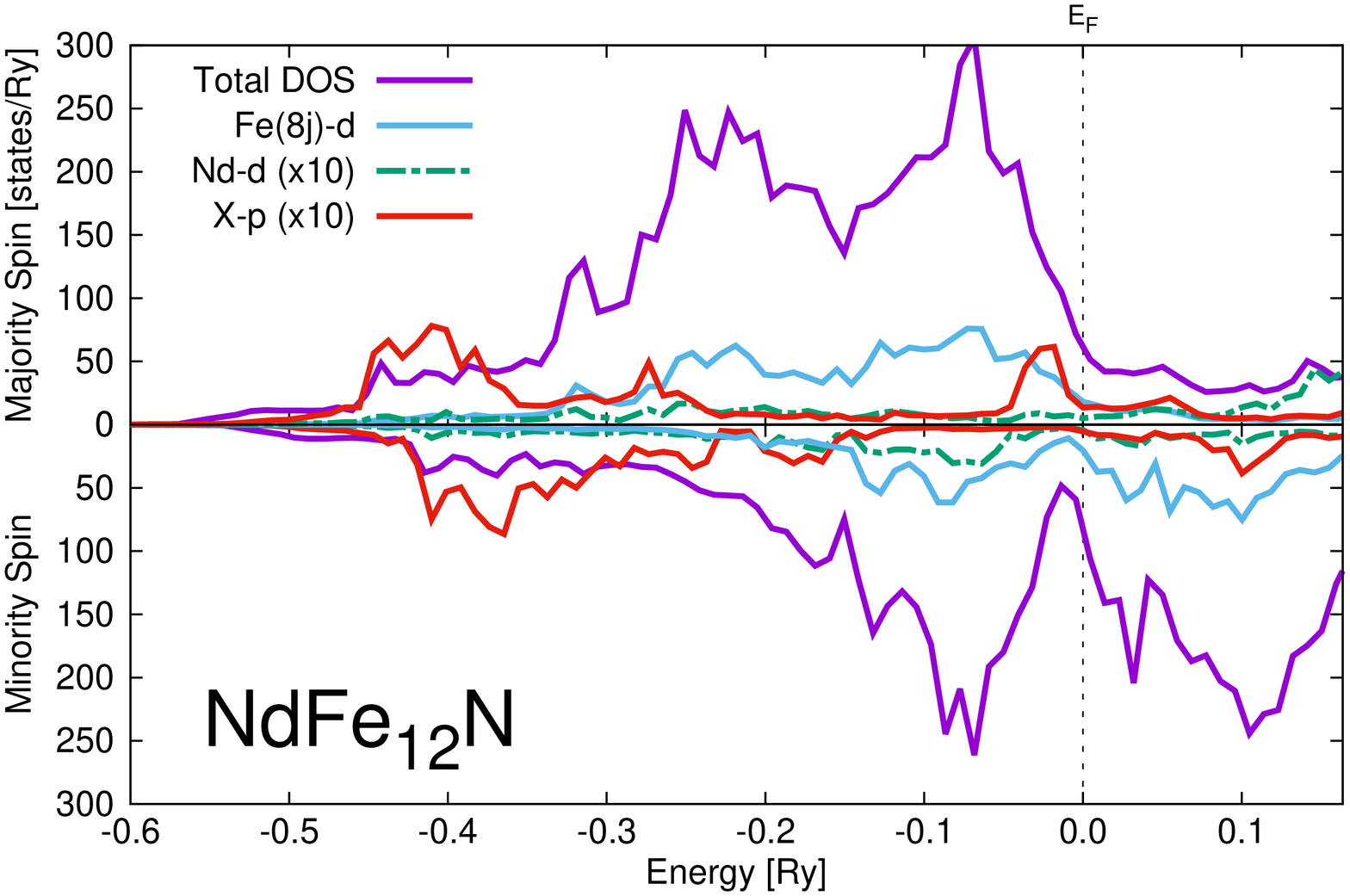}
 \includegraphics[width=5.5cm,angle=270]{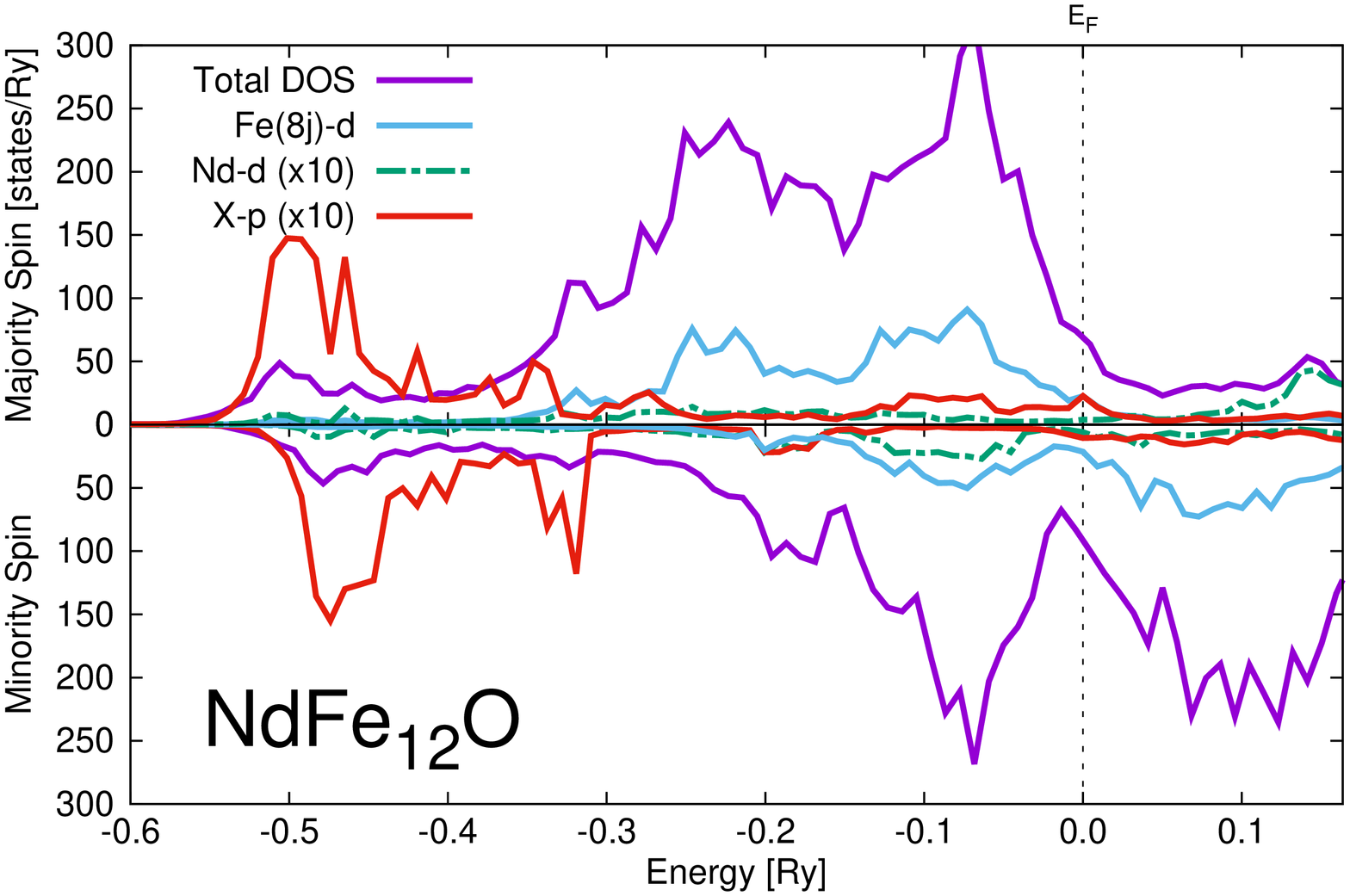}
 \includegraphics[width=5.5cm,angle=270]{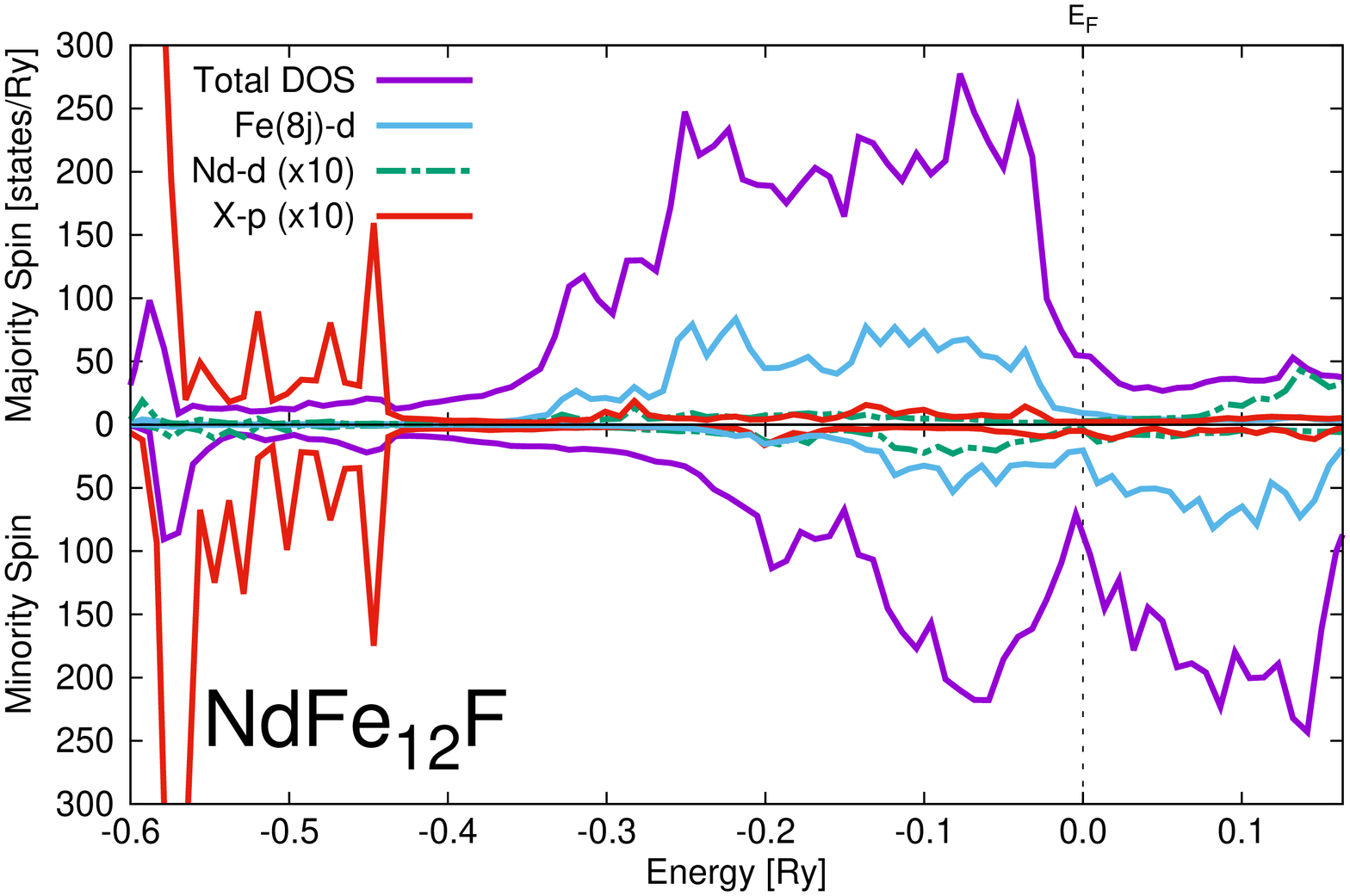}
  \caption{\label{DOS}
(Color online) 
  Total and partial DOS of NdFe$_{12}$ and NdFe$_{12}$X
  (X = B, C, N, O, F).
  Note that the partial density values of the Nd-d states and
  the X-p states are multiplied by 10 to show the detail of the curves.
  }
 \end{figure*}

Figure \ref{JRT_fixed_lattice_DOS} plots the density of states (DOS) at $Z_{\rm X}$ = 6.8. 
The X-p DOS has a peak at the Fermi level in the majority-spin channel, 
but there is no feature in the minority channel at that point.
This type of DOS distribution
has been discussed previously 
\cite{Asano93,Akai95,Harashima15e}
to explain the change in the total spin moment
caused by X.
The discussion is also useful for understanding
the reduction of the Nd moment,
and the results of our calculation also show
good agreement with the accompanying theory.
The X-2p state hybridizes with states of the
neighboring Fe sites. 
An antibonding state between them appears 
above the Fermi level in the cases of X = B and X = C. 
As $Z_{\rm X}$ increases, the X-2p level becomes deeper. 
Consequently, this hybridized state is pulled down (Fig. \ref{DOS})
and crosses the Fermi level at $Z_{\rm X}\sim 7$ in the
majority-spin
channel,
which leads to enhancement of the magnetic moment of the whole compound
(although the magnetovolume effect also enhances the magnetic moment, 
the dependence of the total moment on X comes mainly from the chemical effect
\cite{Harashima15e}).

This hybridized state in the majority-spin channel
has some weight at the Nd site. 
Hence, as the occupation number of this state increases, 
the majority-spin density becomes larger.
This results in a decrease in the magnitude of the local spin moment
at the Nd site
because it is antiparallel to the total spin moment
(electrons in the majority-spin channel of the entire system
are a minority at the Nd site).
This reduction leads to weakening of the spin-rotational
perturbation considered in Liechtenstein's formula
and weakens
$J_{\rm Nd\operatorname{--}Fe(8j)}$.
To summarize, filling of the hybridized state affects both the
magnetization and the Nd--Fe magnetic coupling, but in opposite
ways.
 
\subsection{Chemical effect of X on the Curie temperature}
\label{Tc_section}
\begin{figure*}
 \includegraphics[width=9cm]{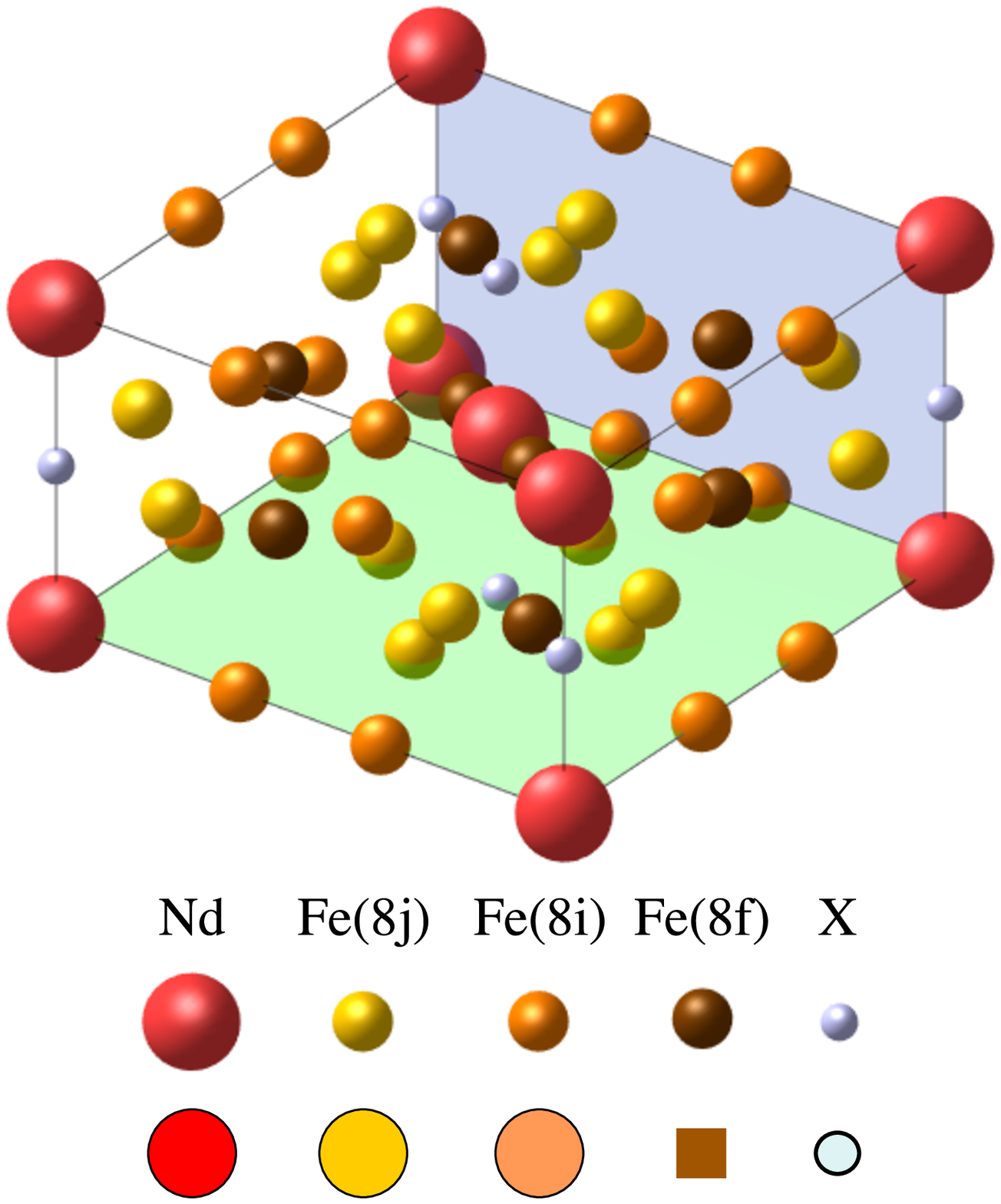}
 \includegraphics[width=5cm]{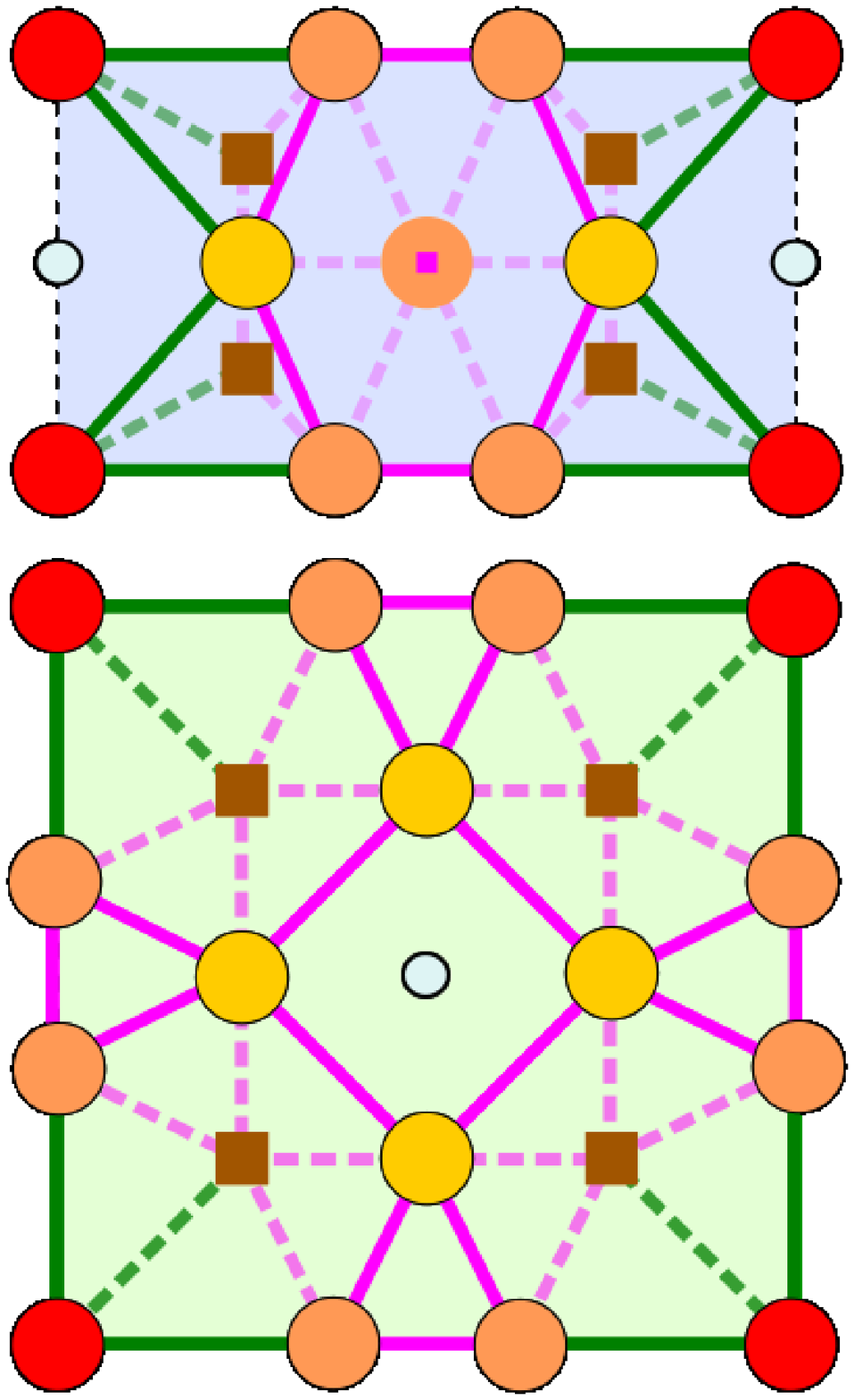}
 \caption{
 (Color online)
 Crystal structure of NdFe$_{12}$X.
 The left figure shows the conventional unit cell of the structure.
 The right figures show cross sections of the cell
 with the ab- and ac-planes indicated by
 light blue and green, respectively, in the left figure.
 The T--T bonds (magenta) and the R--T bonds (green) in the smaller set,
 $\mathcal{S}$, are
 also shown. Solid lines indicate in-plane bonds, and
 dashed lines indicate out-of-plane bonds.
 \label{structure}
}
\end{figure*}

All values of the Curie temperature are
computed within the mean-field approximation
from $J_{i,j}$. 
We use two different sets of $J_{i,j}$ for comparison. 
The larger set, $\mathcal{L}$, is composed of $J_{i,j}$ for the bonds
with lengths shorter than
0.9$a$, where $a$ is the side length of the base square for the
conventional unit cell.
The smaller set, $\mathcal{S}$, is the union of the following two subsets:
(i) 
$J_{i,j}$ for the shortest bonds 
in the classes of the site combinations
(e.g., Fe(8j)--Fe(8i), Fe(8f)--Nd and
X--X);
and
(ii) 
$J_{i,j}$ for the bonds
shorter than
3.80 \AA.
The geometries of the T--T and R--T bonds in $\mathcal{S}$ are
shown in Fig. \ref{structure}.
%
As a result, 
all bonds between the Fe's and 
their 3rd nearest Fe-neighbors are included in ${\mathcal S}$.
When deciding on this cutoff of 3.80 {\AA},
we referred to
a previous Monte Carlo study of Nd$_2$Fe$_{14}$B
(Ref. \onlinecite{Toga16}),
where results with a cutoff of 3.52 {\AA}
are in good agreement with 
results calculated with a cutoff of
15.8 {\AA}.

\begin{figure*}
 \includegraphics[width=5.5cm,angle=270]{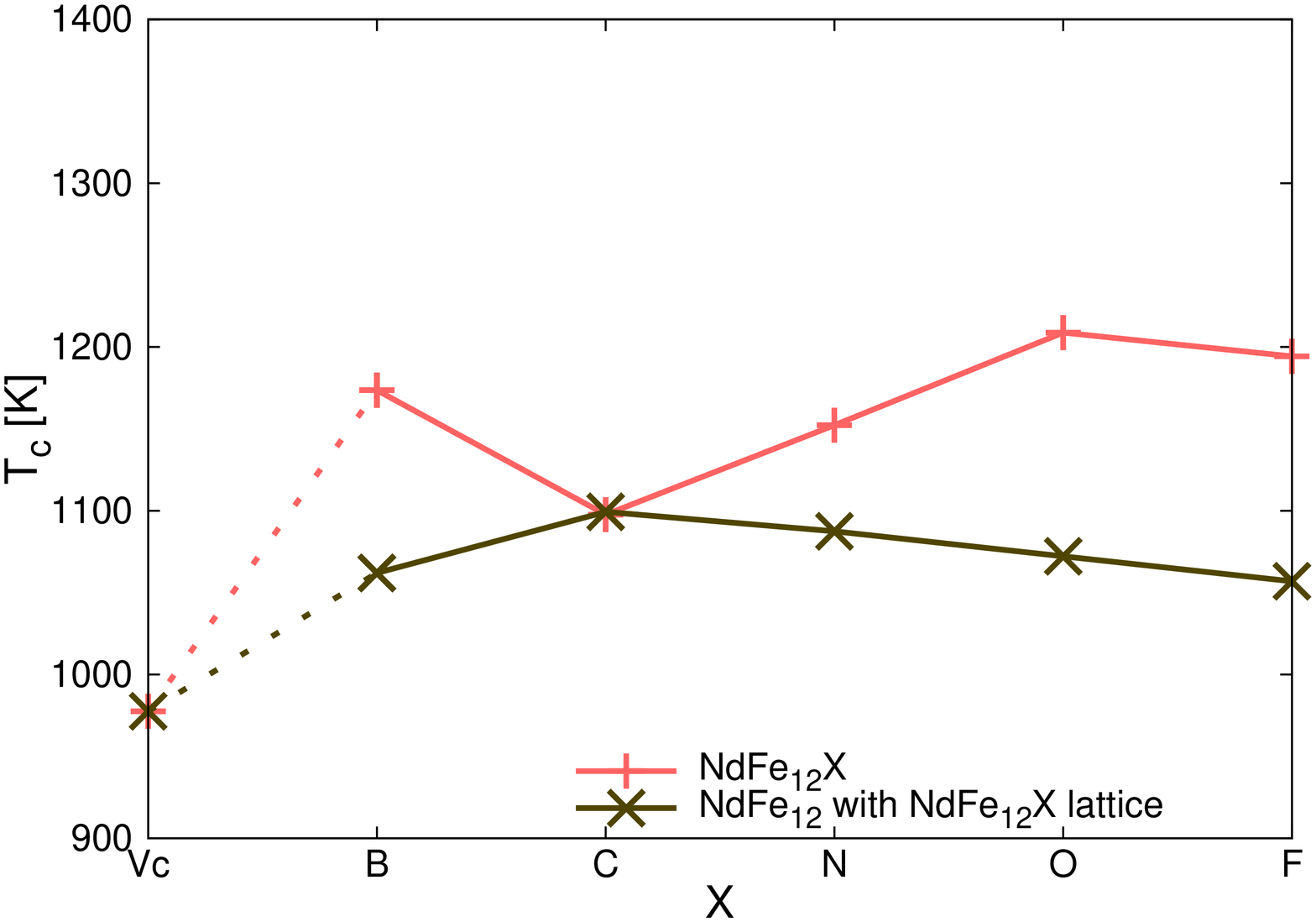}
 \includegraphics[width=5.5cm,angle=270]{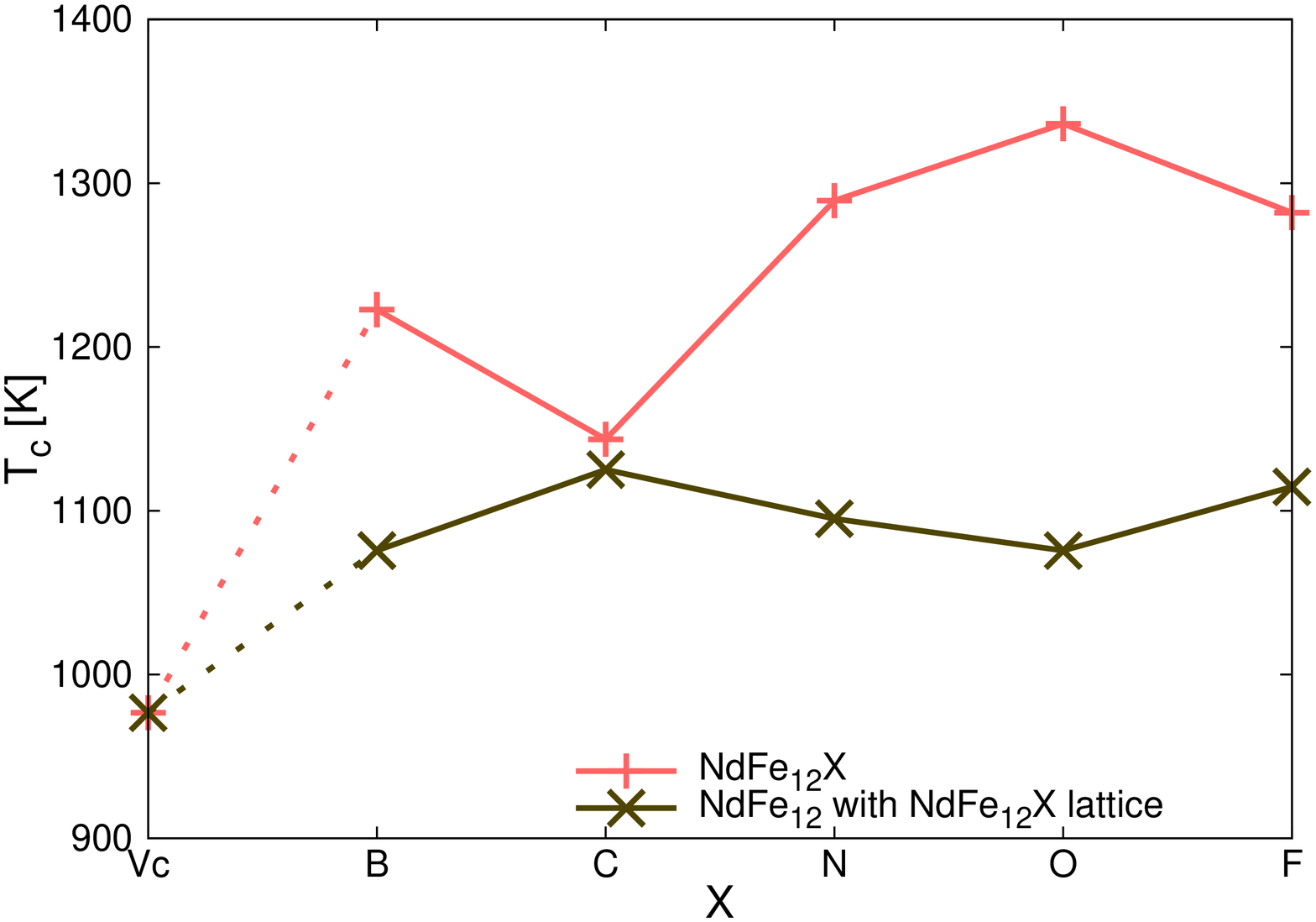}
 \caption{
(Color online) 
 The Curie temperature of NdFe$_{12}$X
 (X = Vc, B, C, N, O, F)
 with the optimized lattices ($+$ symbols)
 and NdFe$_{12}$\#NdFe$_{12}$X ($\times$ symbols).
 NdFe$_{12}$Vc denotes NdFe$_{12}$.
 The data in the left figure are calculated from the larger $J_{i,j}$
 set, $\mathcal{L}$; 
 the data in the right figure are
 calculated from the smaller set, $\mathcal{S}$.
 \label{Tc}
}
\end{figure*}

Figure \ref{Tc} shows the Curie temperatures
of NdFe$_{12}$
and NdFe$_{12}$X 
obtained from the sets $\mathcal{L}$ (left) and 
$\mathcal{S}$ (right). 
Results for NdFe$_{12}$\#NdFe$_{12}$X
are also presented. 
The overall behavior
of the two curves 
is similar between the two panels,
which would justify our use of the smaller set $\mathcal{S}$
in the analysis of $J_{i,j}$ dependence of the Curie temperatures
appearing later.
%
%
The Curie temperature 
is enhanced by the introduction of X in all cases studied. 
Comparing the results for NdFe$_{12}$X with 
NdFe$_{12}$\#NdFe$_{12}$X,
it can be seen that 
the enhancement originates primarily from the structure
as for X = C. 
The same order of enhancement
originating from
the structural
change (mainly the volume expansion)
is also seen
in the cases of X = B, N, O, F.
However, the structural change
accounts only for approximately half of the enhancement.
The chemical effects are as important as
the structural change.
 
\label{SubSecTc}
The Curie temperature
of NdFe$_{12}$N shown in Fig. \ref{Tc} is much
higher than the experimental value of $T_{\rm C} \approx 820\ \mathrm{K}$
obtained by Hirayama et al.~\cite{Hirayama15}.
This overestimation in our calculation presumably comes from the use of
the mean field approximation,
which almost certainly overestimates $T_{\rm C}$ of spin models.
However, the calculated difference between $T_{\rm C}$ before and after the
nitrogenation is comparable to the experimental value of
$T_{\rm C}^\mathrm{NdTiFe_{11}N_x} - T_{\rm C}^\mathrm{NdTiFe_{11}}
\approx 200\ {\mathrm K}$ based on 
$T_{\rm C}^\mathrm{NdTiFe_{11}N_x}$ from Ref. \onlinecite{Yang91} and
$T_{\rm C}^\mathrm{NdTiFe_{11}}$ from Ref. \onlinecite{Hu89}.

To see how the Curie temperature depends on
the change in $J_{i,j}$ caused by the introduction of X,
we calculate the following quantity: 
\begin{equation}
 \Delta T_{\rm C}^{(k,l)}
  =
  T_{\rm C}[\{J^{\rm NdFe_{12}X}_{i,j}\}]
  -
  T_{\rm C}[\{J^{{\rm LOU},(k,l)}_{i,j}\}],
  \label{Def_DeltaTc}
\end{equation}
where $T_{\rm C}[\{J_{i,j}\}]$ is the Curie temperature calculated
from the set of magnetic couplings $\{J_{i,j}\}$, and
$J^{{\rm LOU},(k,l)}_{i,j}$ is defined as follows:
\begin{equation}
 J^{{\rm LOU},(k,l)}_{i,j}
  =
  \left\{
   \begin{array}{cc}
    J_{i,j}^\mathrm{NdFe_{12}} &
     \mbox{[$(i,j)$ is equivalent to $(k,l)$]}\\
    J_{i,j}^\mathrm{NdFe_{12}X} &
     \mbox{(otherwise)}
   \end{array}
  \right.
  ,
  \label{DefLOU}
\end{equation}
where LOU stands for ``leave-one-unchanged.''
Note that this $\Delta T_{\rm C}^{(k,l)}$
is positive when the change in $J_{k,l}$ caused by
the introduction of X
enhances the Curie
temperature.
We also use a similar quantity $\tilde{\Delta} T_{\rm C}^{(k,l)}$
by replacing
$J_{i,j}$ in 
the reference NdFe$_{12}$ system
in equation \eqref{DefLOU}
with the values from NdFe$_{12}$\#NdFe$_{12}$X.
In this analysis, we focus on the bonds in $\mathcal{S}$
and use only their $J_{i,j}$ values
to calculate $\Delta T^{(k,l)}_{\rm C}$.

\begin{figure*}
 \includegraphics[width=5.5cm,angle=270]{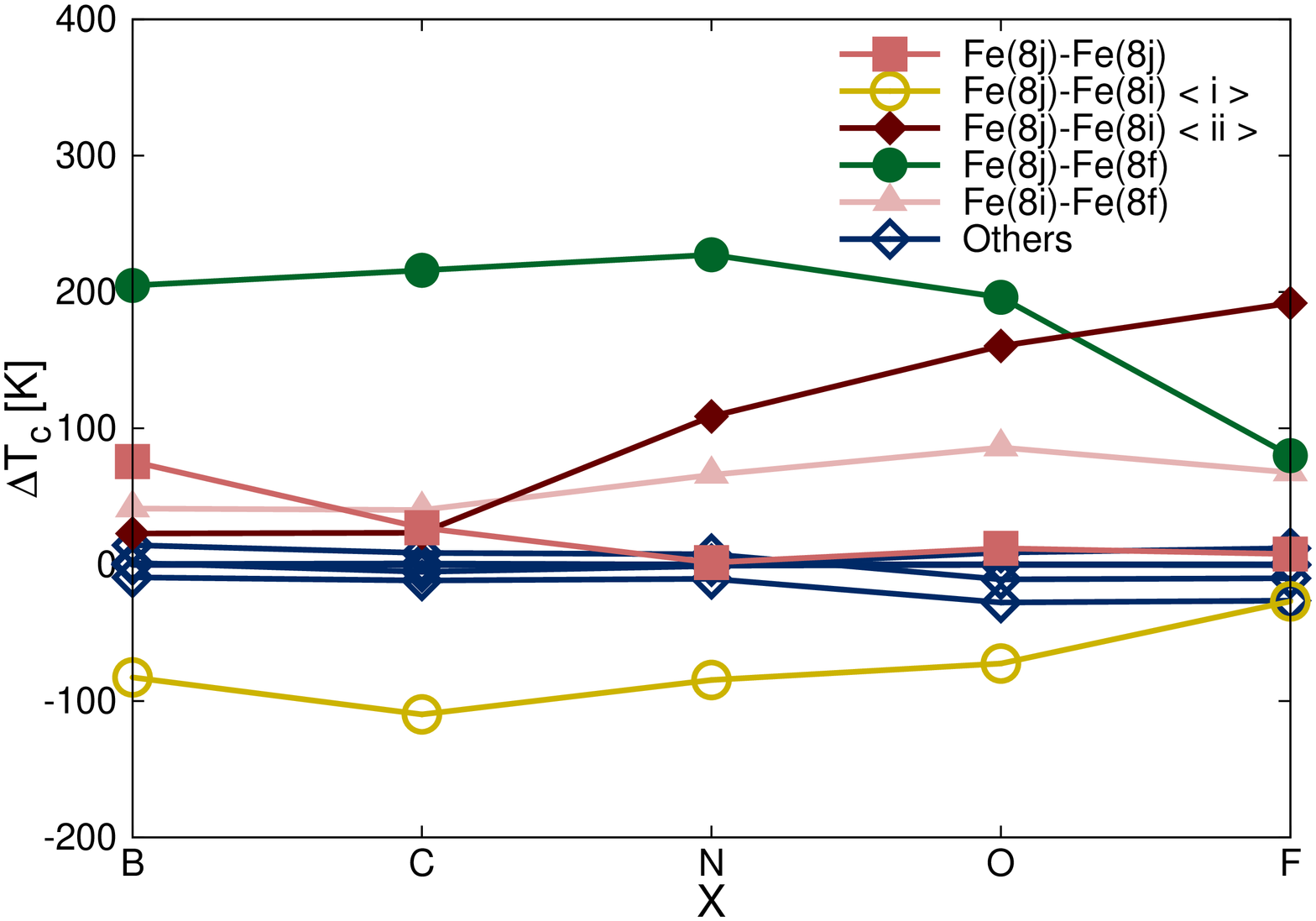}
 \includegraphics[width=5.5cm,angle=270]{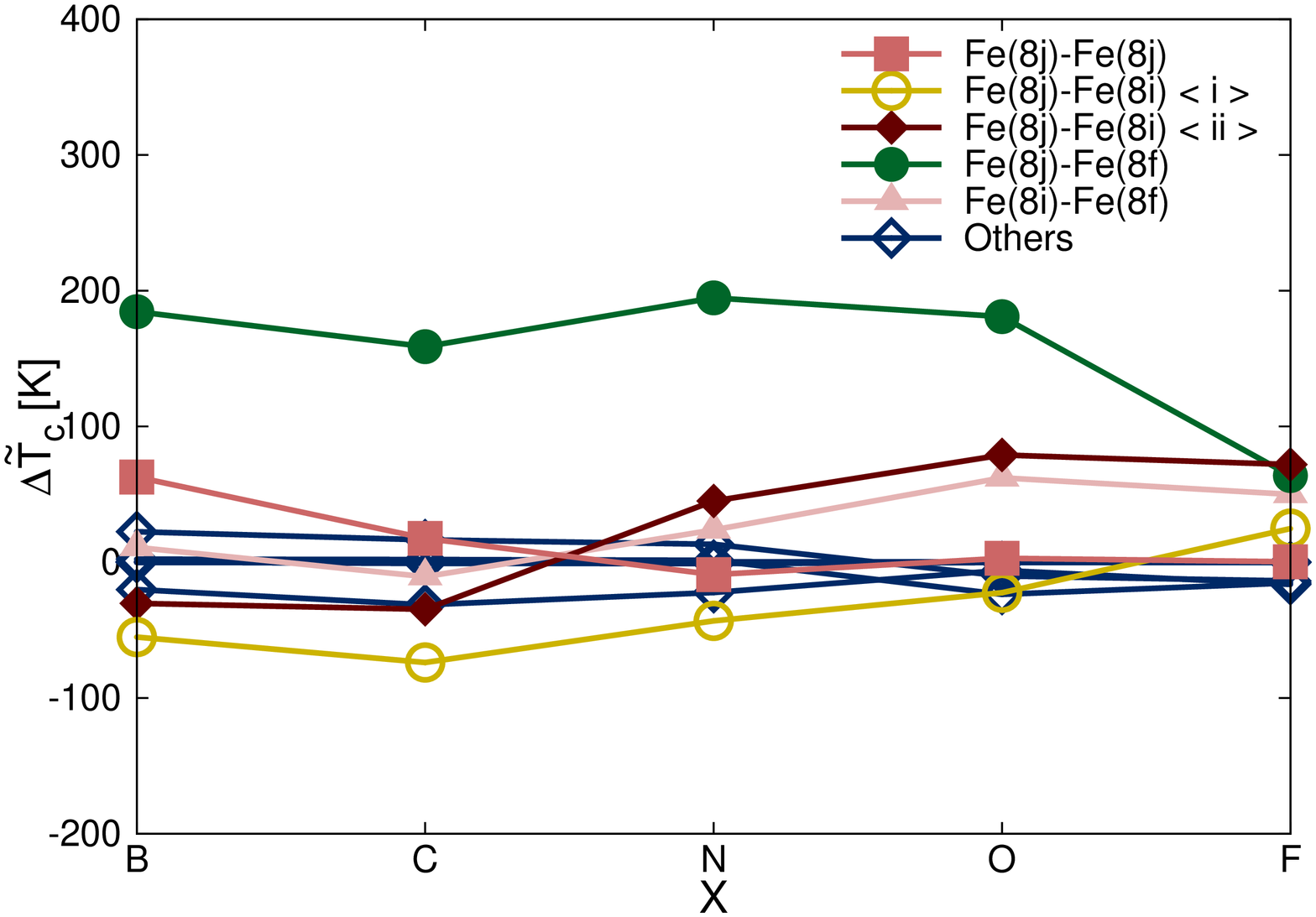}
 \caption{
(Color online) 
 (Left)
 Difference in Curie temperature
 $\Delta T_{\rm C}^{(k,l)}$
as defined by equation \eqref{Def_DeltaTc},
 and (Right)
 $\tilde{\Delta} T_{\rm C}^{(k,l)}$ in which the lattice parameters
 of the
 reference NdFe$_{12}$ system [see the first line in Eq.~\eqref{DefLOU}]
 are set to the parameters of
 NdFe$_{12}$X.
  \label{DeltaTc}
}
\end{figure*}

The left panel in Fig. \ref{DeltaTc} shows
$\Delta T_{\rm C}^{(k,l)}$
as a function of X. 
There are two Fe(8i)--Fe(8i) bonds and two Fe(8j)--Fe(8i) bonds 
in the $\mathcal{S}$ subset.
To distinguish one from the other,
we denote the shortest bonds by $<$i$>$
and the second-shortest bonds by $<$ii$>$ 
in Fig. \ref{DeltaTc} and the following discussion.
From Fig. \ref{DeltaTc}, it can be seen that
the values of $\Delta T_{\rm C}^{\rm T\operatorname{--}T}$
($\Delta T_{\rm C}^{(k,l)}$ for T--T bonds)
are much larger than 
the values of $\Delta T_{\rm C}^{\rm R\operatorname{--}T}$ and
$\Delta T_{\rm C}^{\rm R\operatorname{--}R}$,
which are denoted by ``Others'' in the figure.
This indicates that changes in the T--T bonds
predominantly cause the enhancement of $T_{\rm C}$. 
The right panel in Fig. \ref{DeltaTc} shows
$\tilde{\Delta} T_{\rm C}^{(k,l)}$. 
We do not find any significant differences between
$\Delta T_{\rm C}^{(k,l)}$
and $\tilde{\Delta} T_{\rm C}^{(k,l)}$, 
which indicates that
the chemical effects significantly contribute to
$\Delta T_{\rm C}^{(k,l)}$,
whereas the structure change plays a minor role.

Both $\Delta T_{\rm C}^{\rm Fe(8j)\operatorname{--}Fe(8f)}$
and
$\tilde{\Delta} T_{\rm C}^{\rm Fe(8j)\operatorname{--}Fe(8f)}$
have the largest magnitude
in the range of X = B--O.
This implies that 
the change of $J_{\rm Fe(8j)\operatorname{--}Fe(8f)}$
caused by the introduction of X has a strong positive effect on
the enhancement of the Curie temperature.
It is noteworthy that even at X = C, where the lattice expansion alone
seems sufficient to explain
the enhancement of the Curie temperature, 
$\tilde{\Delta}T_{\rm C}^{\rm Fe(8j)\operatorname{--}Fe(8f)}$
is large and not very different from that at X = B, N, O.
Therefore, 
whereas the change in $J_{\rm Fe(8j)\operatorname{--}Fe(8f)}$ gives the largest contribution, 
it does not explain
the dependence of the Curie temperature on X. 
In the case of X = B, the difference from X = C mainly comes from
the increase in 
$\tilde{\Delta}T_{\rm C}^{\rm Fe(8j)\operatorname{--}Fe(8j)}$, whereas
in the case of X = N, O, F, it comes mainly from
the increase in
$\tilde{\Delta}T_{\rm C}^{\rm Fe(8j)\operatorname{--}Fe(8i)<ii>}$ and
$\tilde{\Delta}T_{\rm C}^{\rm Fe(8i)\operatorname{--}Fe(8f)}$.
This leads us to believe that the mechanism behind the enhancement of
the Curie temperature differs between the case of X = B and those of
X = N, O, F.

\section{Conclusion}
\label{conclusion}
We studied and investigated the internal magnetic couplings of NdFe$_{12}$ and NdFe$_{12}$X for X = B, C, N, O, F by first-principles calculations
and 
found that the introduction of nitrogen
to NdFe$_{12}$ reduces the strength of R--T magnetic couplings
owing to Nd--X hybridization,
with $J_{\rm Nd\operatorname{--}Fe(8j)}$ particularly
reduced so significantly that lattice expansion due to the nitrogen
cannot compensate for the reduction.
Although nitrogen is often used to enhance
the magnetic properties of magnetic compounds,
our results suggest that nitrogenation
may have countereffects on the anisotropy field
of NdFe$_{12}$
at finite temperatures.

We also evaluated the Curie temperatures
of NdFe$_{12}$ and NdFe$_{12}$X
within the mean field approximation
and found that the volume expansion caused by the introduction of X
cannot explain all enhancement of $T_{\rm C}$.
The introduction of X causes significant changes in the magnetic couplings
of NdFe$_{12}$ and has a significant effect on the Curie temperature.
Nitrogen was found to enhance the Curie temperature, 
as found experimentally in similar compounds. 
Oxygen and fluorine were also found to enhance $T_{\rm C}$ as much as nitrogen. 
Although boron also produced the same order of positive effect
on $T_{\rm C}$
within the framework above
(see also Appendix \ref{LMD_section} for results with a model
for the paramagnetic state),
the mechanism appears to be different from that
for the cases of X = N, O, F.

\begin{acknowledgments}
 The authors are grateful for support from
 the Elements Strategy Initiative Project under the auspices
 of MEXT.
 This work was also supported by
MEXT as a social and scientific priority issue (Creation of new functional Devices and high-performance Materials to 
Support next-generation Industries; CDMSI) to be tackled by using the post-K computer.
The computation was partly carried out using the facilities of 
the Supercomputer Center, the Institute for Solid State Physics,
the University of Tokyo, and the supercomputer of ACCMS, Kyoto University.
This research also used computational resources of the K computer provided by the RIKEN Advanced Institute for Computational Science through the HPCI System Research project (Project ID:hp170100).
%
\end{acknowledgments}

\appendix
\section{Lattice parameters}
\label{lattparams}
Table \ref{table2} shows the optimized lattice parameters for
NdFe$_{12}$ and NdFe$_{12}$X (X = B, C, N, O, F)
that we used in our calculations.
The parameters $p_{\rm 8i}$ and $p_{\rm 8j}$ correspond to
the atomic positions described in table \ref{table1}.
	
\begin{table}[h]
  \caption{
 Optimized lattice parameters for NdFe$_{12}$X
 (X = Vc, B, C, N, O, F), where
 NdFe$_{12}$Vc
 denotes NdFe$_{12}$.
 For the definitions of the inner parameters $p_{\rm 8j}$
 and $p_{\rm 8i}$, see table \ref{table1}.
  \label{table2}}
 \begin{tabular}{ccccc}
  \hline
  \hline
   X & $a$ [\AA] & $c$ [\AA] &$p_{\rm 8i}$& $p_{\rm 8j}$\\
  \hline
                         Vc &  8.533 &   4.681 & 0.3594 & 0.2676 \\
                          B &  8.490 &   4.933 & 0.3599 & 0.2683 \\
                          C &  8.480 &   4.925 & 0.3606 & 0.2756 \\
                          N &  8.521 &   4.883 & 0.3612 & 0.2742 \\
                          O &  8.622 &   4.794 & 0.3608 & 0.2670 \\
                          F &  8.782 &   4.720 & 0.3594 & 0.2487 \\
  \hline
  \hline
 \end{tabular}
	\end{table}

\begin{table}[h]
 \caption{Atomic positions of the elements assumed in our calculation
 for
 NdFe$_{12}$ and NdFe$_{12}$X (X = B, C, N, O, F).
 The variables, $x$, $y$, and $z$
 denote the point ($ax$, $ay$, $cz$) in Cartesian coordinates.
 \label{table1}
 }
 \begin{tabular}{ccrrr}
  \hline
  \hline
  Element & Site & $x$ & $y$& $z$\\
  \hline
  Nd & 2a &            0 &    0 &    0 \\
  Fe & 8f &         0.25 & 0.25 & 0.25 \\
  Fe & 8i & $p_{\rm 8i}$ &    0 &    0 \\
  Fe & 8j & $p_{\rm 8j}$ &  0.5 &    0 \\
   X & 2b &            0 &    0 &  0.5 \\
  \hline
  \hline
 \end{tabular}
 \end{table}

\section{Total and local moments}
\label{moments}
\subsection{Comparison with full-potential calculation}

\begin{figure}
 \includegraphics[width=6cm,angle=270]{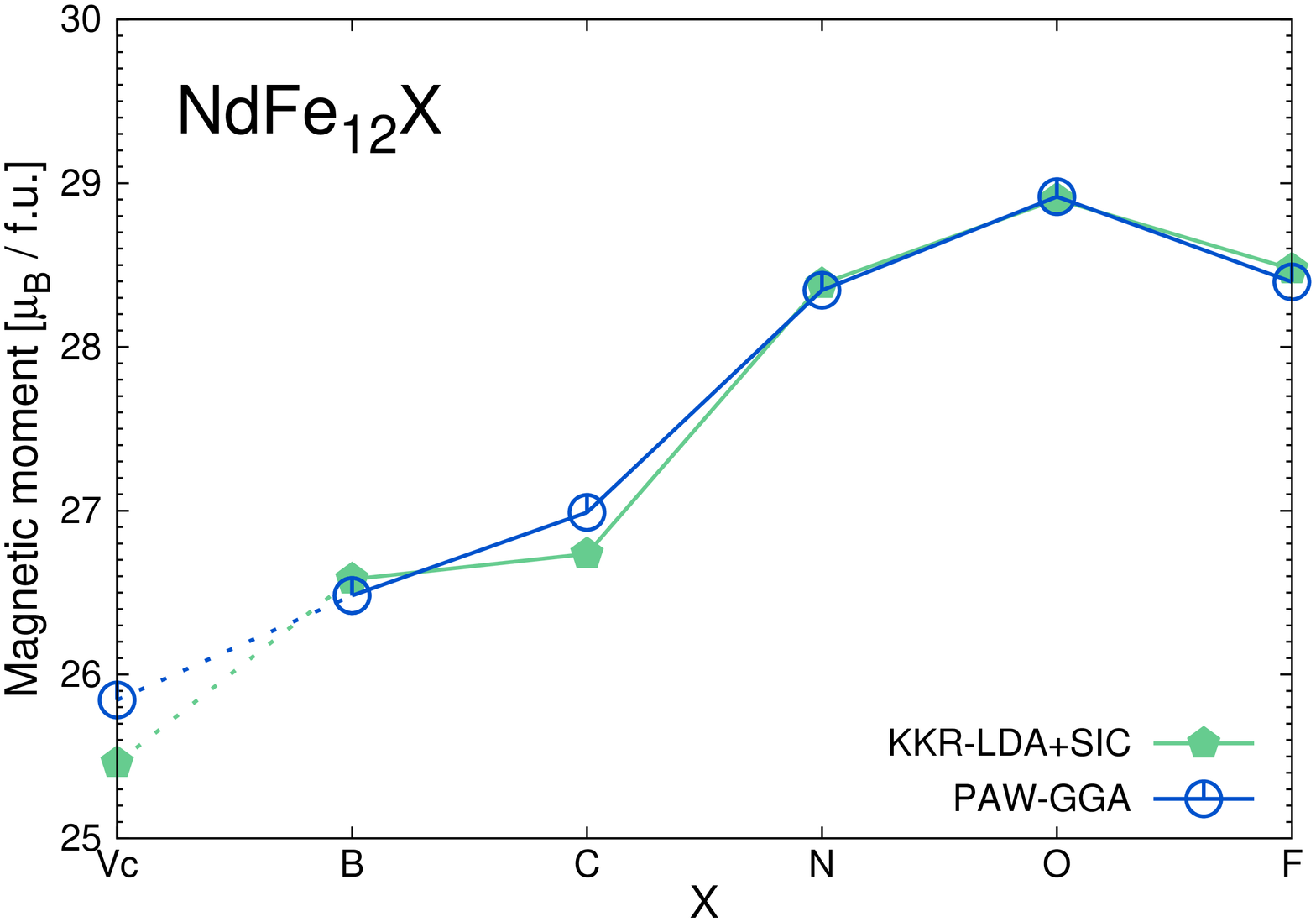}
 \caption{(Color online) \label{total_moment}
 Total magnetic moment
 of NdFe$_{12}$X
 (X = Vc, B, C, N, O, F),
 where 
 NdFe$_{12}$Vc denotes NdFe$_{12}$.
}
\end{figure}

\begin{figure*}
 \includegraphics[width=5cm,angle=270]{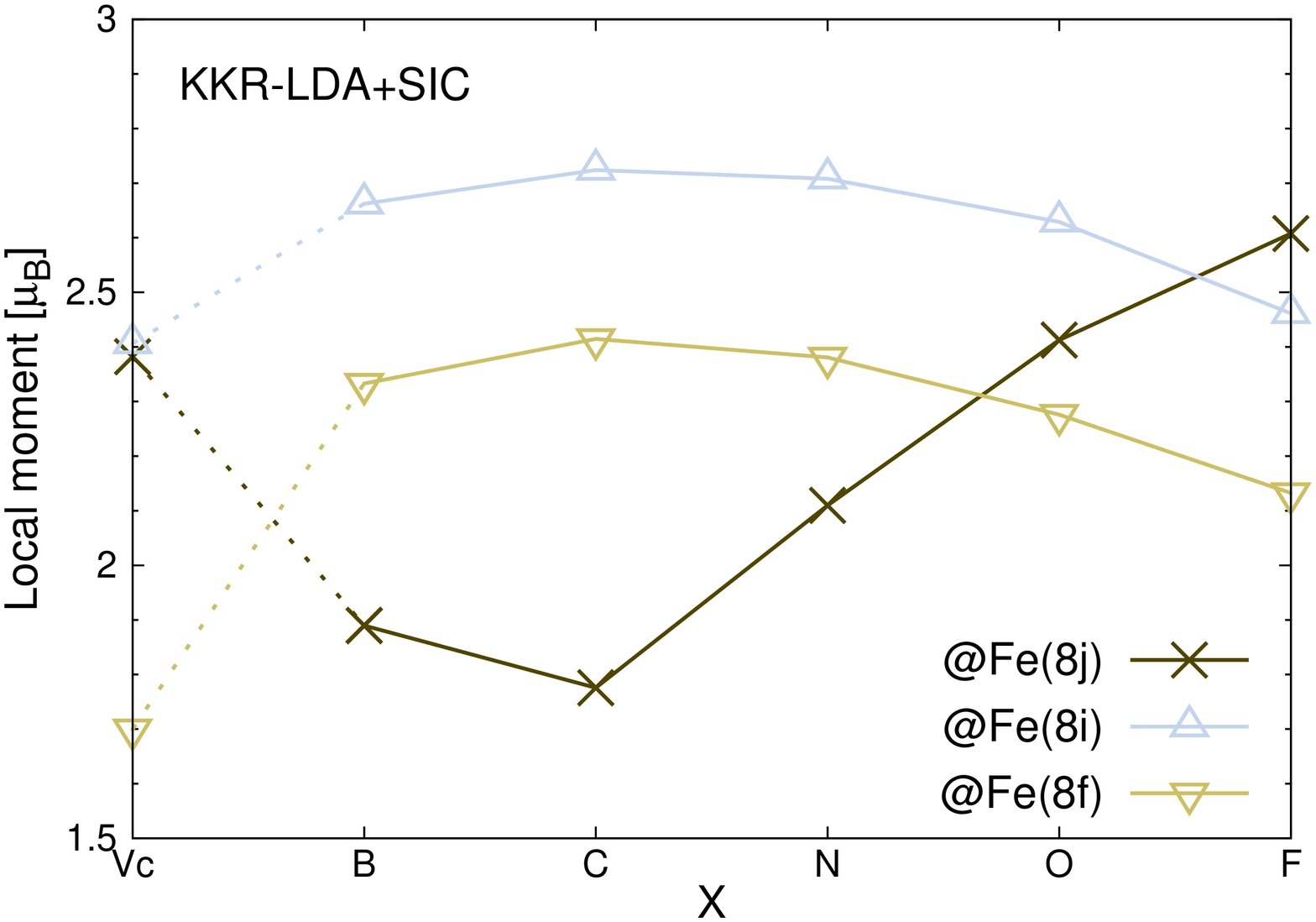}
 \includegraphics[width=5cm,angle=270]{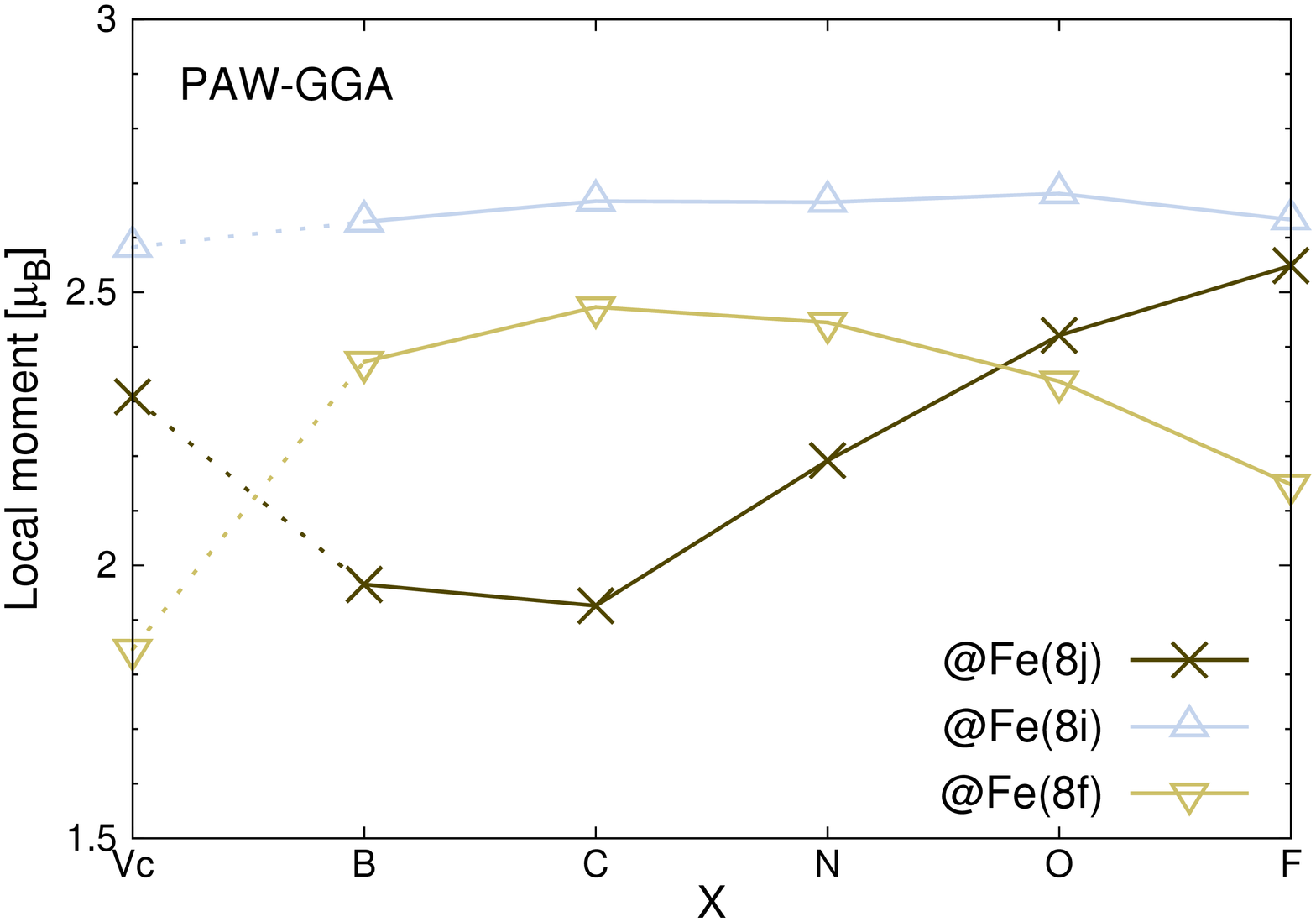}
 \includegraphics[width=5cm,angle=270]{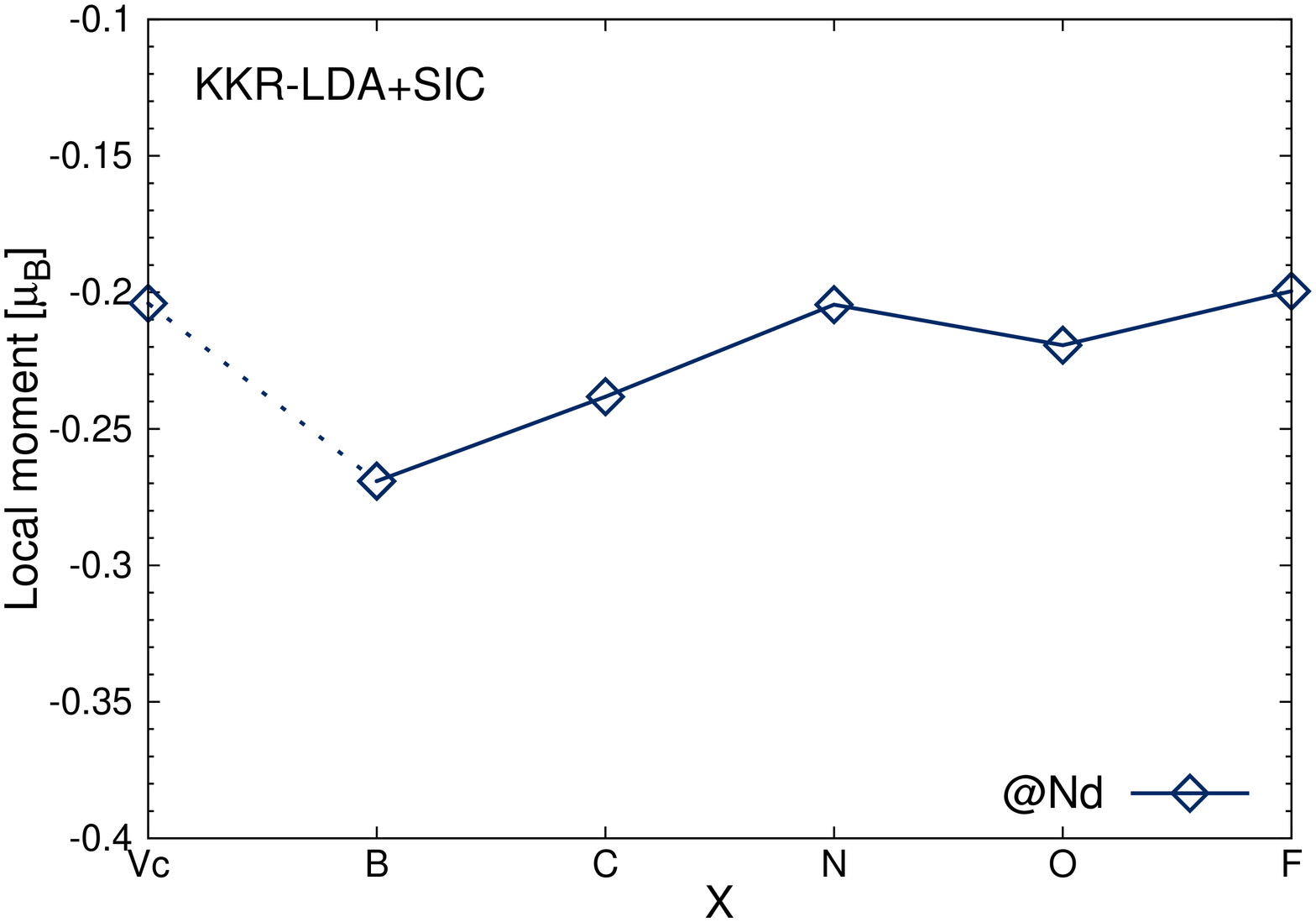}
 \includegraphics[width=5cm,angle=270]{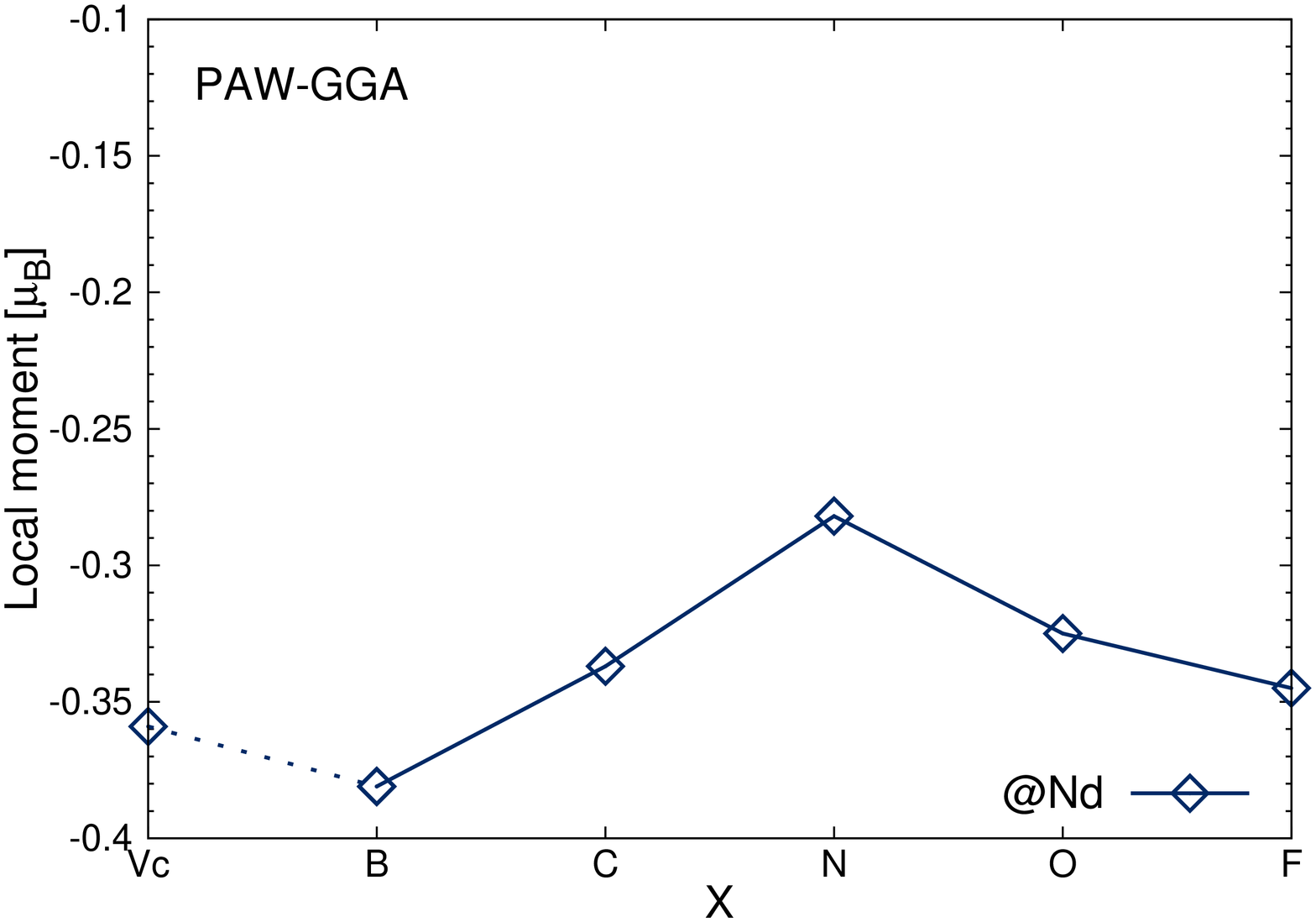}
 \includegraphics[width=5cm,angle=270]{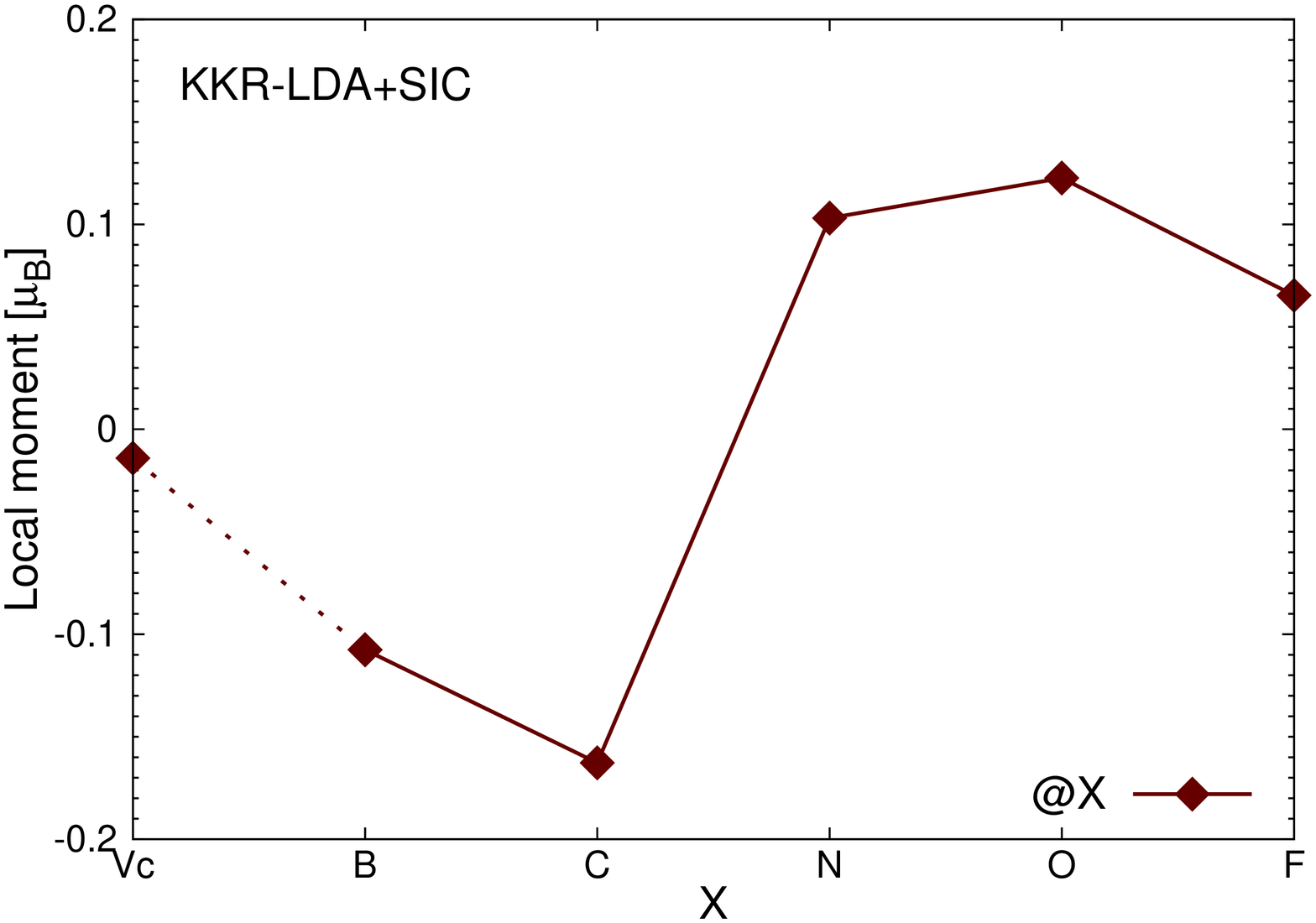}
 \includegraphics[width=5cm,angle=270]{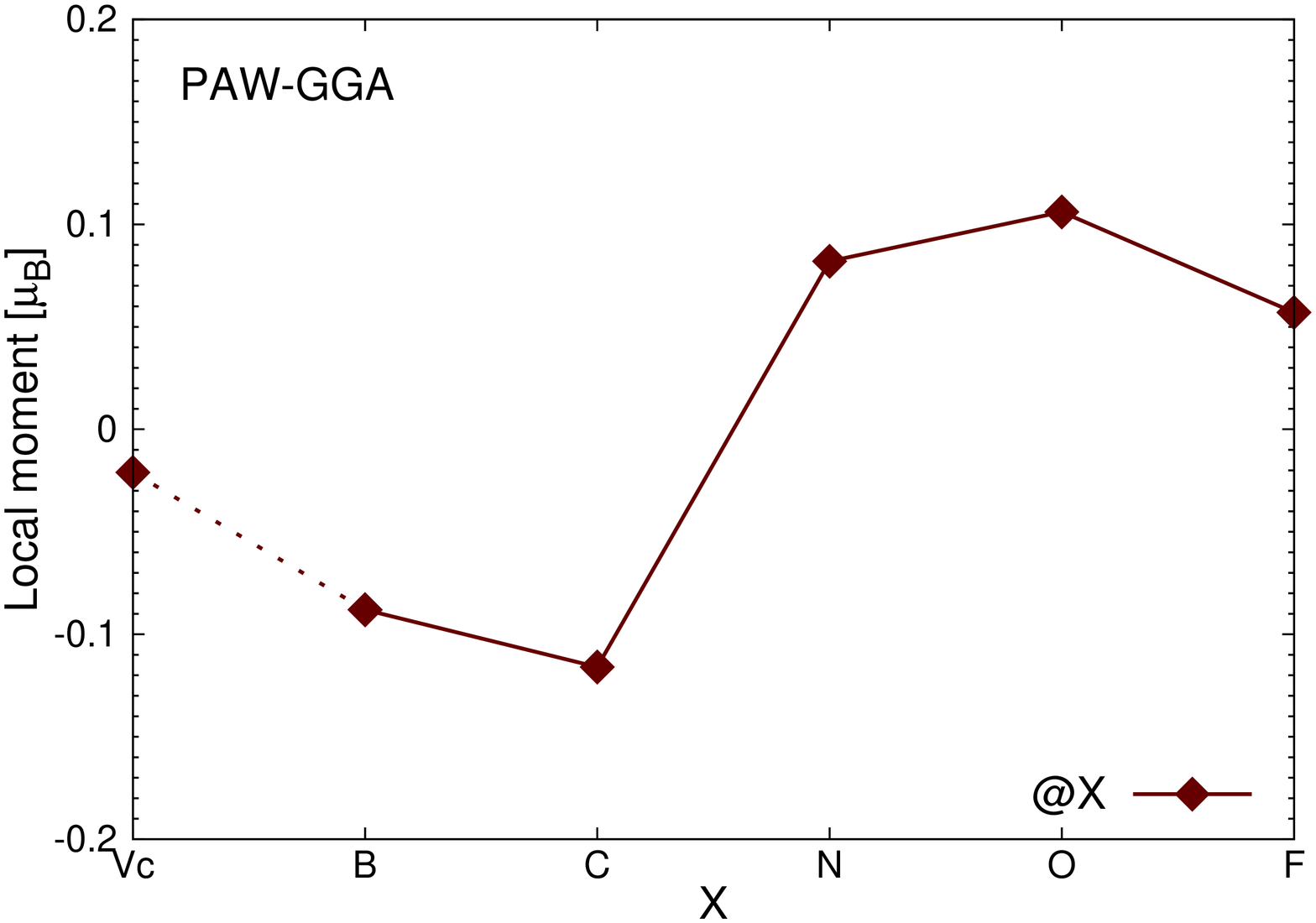}
 \caption{(Color online) \label{local_moment}
 Local magnetic moments at the Fe sites (top),
 the Nd site (middle), and the X site (bottom)
 in NdFe$_{12}$X  (X = Vc, B, C, N, O, F),
 where NdFe$_{12}$Vc denotes NdFe$_{12}$.
 The data in the left figures are from the KKR-LDA+SIC calculation;
 the data in the right figures are from the PAW-GGA calculation
 with the local moments defined as integrated spin density within
 the muffin-tin radii used in the KKR calculation.
}
\end{figure*}

Figure \ref{total_moment} shows the total moment of
NdFe$_{12}$ and NdFe$_{12}$X (X = B, C, N, O, F);
Fig. \ref{local_moment} shows the local moments of
NdFe$_{12}$ and NdFe$_{12}$X (X = B, C, N, O, F).
In both, the values from KKR-LDA+SIC are 
compared with those from the full-potential calculation with 
PAW-GGA.
In both cases, the regions of integration to obtain
the local moments are set to
the spheres with the muffin-tin radii used in the KKR calculation.
The contribution from Nd-f orbitals are excluded in those plots
as mentioned in section \ref{framework} of the main text.

\subsection{Local moment disorder}
\label{LMD_section}
\begin{figure}
 \includegraphics[width=6cm,angle=270]{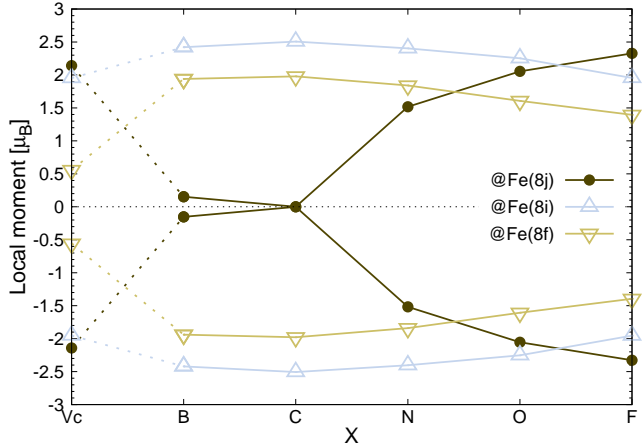}
 \caption{(Color online) \label{LMD_local_moment}
 Values of the local magnetic moments at the Fe sites
 for
 NdFe$_{12}$ (at X = Vc) and
 NdFe$_{12}$X (X = B, C, N, O, F)
 in the state of local moment disorder\cite{Akai93}.
}
\end{figure}

We also performed calculations for NdFe$_{12}$ and 
NdFe$_{12}$X with the local moment disorder (LMD)
model\cite{Akai93}
to approximate the paramagnetic states.
In the calculation, each of the atomic sites,
$\mathcal{A}$ (=Nd, Fe, X),
is described by twofold atomic potentials, $\mathcal{A}^\uparrow$ and 
$\mathcal{A}^\downarrow$, where $\mathcal{A}^\uparrow$
has opposite spin-polarization
to $\mathcal{A}^\downarrow$, and they are treated as 
potentials of distinct atoms
that can occupy the $\mathcal{A}$ site with 50\% probability.
This randomness is treated with the coherent potential approximation.
In this hypothetical
(Nd$^\uparrow_{0.5}$Nd$^\downarrow_{0.5}$)
(Fe$^\uparrow_{0.5}$Fe$^\downarrow_{0.5}$)$_{12}$
  X$^\uparrow_{0.5}$ X$^\downarrow_{0.5}$
system, the absolute value of the Fe(8j) moment is greatly reduced
for X = B and C  as shown in Fig. \ref{LMD_local_moment}.
In contrast, for X = N, O, F, the reduction is  much smaller.

\begin{figure}
 \includegraphics[width=6cm,angle=270]{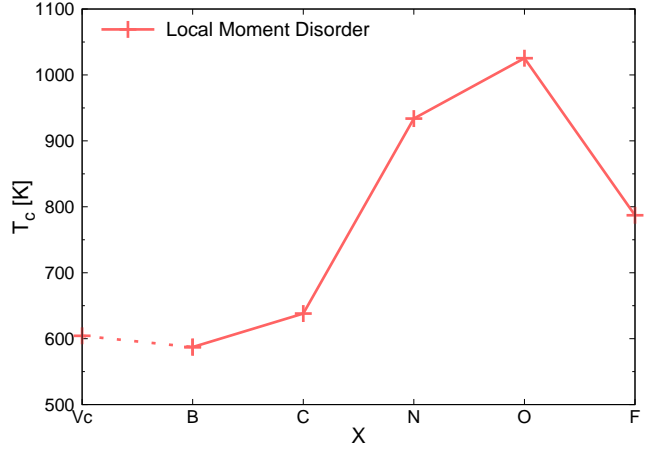}
 \caption{(Color online) \label{LMD_Tc}
 Values of the Curie temperature for
 NdFe$_{12}$ (at X = Vc) and
 NdFe$_{12}$X (X = B, C, N, O, F)
 in the state of local moment disorder \cite{Akai93}.
}
\end{figure}

Intersite magnetic couplings in this system can be compared with
$J_{i, j}$ in the main text
by using Liechtenstein's $J_{i, j}$ between
$\mathcal{A}^\uparrow_i$ and $\mathcal{A}^\uparrow_j$
embedded in this model system.
To compare them in terms of temperature,
we show the Curie temperature for 
NdFe$_{12}$ and NdFe$_{12}$X calculated 
from the thus defined $J_{i, j}$ in Fig. \ref{LMD_Tc}.
The cutoff bond length for this $J_{i, j}$ is 0.9$a$, which is identical to
that for $\mathcal{L}$ in Section \ref{Tc_section}.
Whereas the fluctuation at the Fe(8j) site seems to offset the enhancement of
magnetism in the case of X = B, C, the exchange interaction in 
X = N--F and Vc (NdFe$_{12}$) seems robust.

\bibliography{mu}
\end{document}